\documentclass[pdflatex,sn-mathphys,iicol]{sn-jnl}

\jyear{2021}%

\theoremstyle{thmstyleone}%
\theoremstyle{thmstyletwo}%

\theoremstyle{thmstylethree}%

\begin{document}

\title[Orbital maneuvers around Titania]{Orbital maneuvers for a space probe around Titania}


\author*[1]{\fnm{Jadilene} \sur{Xavier}}\email{jadilene.rodrigues@unesp.br}

\author[2]{\fnm{Antônio Bertachini} \sur{A. Prado}}\email{antonio.prado@inpe.br}
\equalcont{These authors contributed equally to this work.}

\author[1]{\fnm{Silvia Giuliatti} \sur{Winter}}\email{giuliatti.winter@unesp.br}
\equalcont{These authors contributed equally to this work.}

\author[1]{\fnm{Andre} \sur{Amarante}}\email{andre.amarante@unesp.br}
\equalcont{These authors contributed equally to this work.}

\affil*[1]{Grupo de Din\^amica Orbital \& Planetologia, São Paulo State University (Unesp), School of Engineering and Sciences, Guaratinguet\'a, CEP 12516-410, S\~ao Paulo, Brazil}

\affil[2]{National Institute for Space Research, Av. dos Astronautas, 1758, S\~ao Jos\'e dos Campos, SP,  Brazil. Professor, Academy of Engineering, RUDN University, Miklukho-Maklaya street 6, Moscow, Russia, 117198}



\abstract{For most space missions, it is interesting that the probe remains for a considerable time around the mission target. The longer the lifetime of a mission, the greater the chances of collecting information
about the orbited body. In this work, we present orbital maneuvers that aim to show how to avoid a collision of a space probe with the surface of Titania. Through an expansion of the gravitational potential to the second order, the asymmetry of the gravitational field due to the coefficient $C_{22}$ of Titania, the zonal coefficient $J_2$, and the gravitational perturbation of Uranus are considered. Two models of coplanar bi-impulse maneuvers are presented. The first maneuver consists of transferring an initial elliptical orbit to a final circular orbit, and the second has the objective of transferring an initial elliptical orbit to a final orbit that is also elliptical. The lag in the inclination and semi-major axis of the orbits is investigated before performing the maneuvers. To point out the best scenarios for carrying out the maneuvers, a study is presented for different points of an orbit where transfers could be made. In addition, a maneuver strategy is presented to correct the variation of the periapsis argument. The results show that maneuvers performed a few days after integration are more economical than maneuvers performed later, a few days before the collision. The economy of the maneuvers is also demonstrated through an analysis of the ratio of the increase in speed to the lifetime.}




\maketitle

\section{Introduction}
\label{intro}

In recent years there has been an increasing interest in sending space missions to the gaseous planets of the solar system. According to \cite{bib1}, NASA and ESA are planning possible missions to the ice giants from 2024 to 2037. These possible missions can help to understand the origin and evolution of these systems. In addition, some missions are directed towards some specific natural satellites, such as the JUpiter ICy moons Explorer (JUICE) mission, launched in 2023, which will investigate Jupiter and its moons, Ganymede, Europa, and Callisto \cite{Grasset2013}.  

Several works look for better orbits around natural satellites. The purpose of these works is to provide essential data to help future space missions that target these satellites. However, sending a probe into space is a process that demands a lot of time and investment. It is desirable that, during a mission, the probe remains in orbit around the observed body for a long period of time. Thus, searching for long-duration orbits and avoiding collisions with the mission target is very important to achieve success during the observation of the desired object.

The work by \cite{Xavier2022} presents a study on orbits around Titania, one of the moons of Uranus, an ice-giant planet. The authors investigate orbits for a space probe around Titania with various eccentricities values. Perturbations due to the third body and the gravitational coefficients of Titania are considered, and through lifetime maps, orbits with long lifetimes are found. The results of numerical simulations showed that orbits with the longest lifetimes were those whose eccentricity is equal to $10^{-3}$. The authors also show that values other than zero for the argument of periapsis and the longitude of the ascending node can prolong the probe's lifetime. Furthermore, a study on lifetime sensitivity caused by possible errors in Titania's gravitational coefficients concluded that orbits with eccentricity equal to $10^{-3}$ are the most affected by these potential errors.

Long-duration orbits are also studied in \cite{Thamis2022}. In this case, the natural satellite is Io, one of Jupiter's Galilean satellites. The authors were able to detect collisions and possible escapes of the probe. Orbits with significant lifetimes ranged from 6 months to 2.3 years. In addition, the study showed that orbits around Io are more susceptible to perturbations due to their short distance from the perturber. Finally, they emphasized the importance of assigning non-zero values for the argument of periapsis and the longitude of the ascending node.

The work developed by \cite{Cinelli2022} presents a study of the lifetime of a probe in a low-altitude and highly inclined orbit around the natural satellite Europa. The results showed that the obliquity of the third body and the nodal phase affect the lifetime of the probe.

Studying long-duration orbits around natural satellites also requires finding how to avoid a probe collision with the proposed mission target. In this sense, in \cite{Cinelli2019}, corrective orbital maneuvers are proposed for a spacecraft around Europa. The results showed that the lifetime can be extended through appropriate strategies of corrective maneuvers. Orbital maneuvers are also investigated in \cite{Ferreira2022}. After analyzing long-term orbits around Saturn's natural satellite, Titan, the authors present the best results for such maneuvers to be carried out.  

Given the above mentioned, in this paper we present models for orbital maneuvers to avoid collision of a probe with the surface of Titania. The first maneuver proposes to transfer the spacecraft from an initial elliptical orbit to a final circular orbit. In the second model, the transfer is from an initial elliptical orbit to a final elliptical orbit, both of which are coplanar. For this second model, we performed maneuvers at various points of the orbit to analyze if it is more feasible for a maneuver to take place close to the collision or within a few days of the mission's lifetime. We also present a maneuver model for correcting small variations of the argument of periapsis. These maneuvers were performed considering points where the deviation of the semi-major axis and the eccentricity are small.    

\section{Mathematical Model}
\label{sec:model}
The system used here includes Titania as the central body, a space probe in orbit around Titania, and Uranus as the third body. Uranus is assumed to be in a Keplerian orbit around Titania with a radius of $ 25.362 \times 10^{3}$~km and a mass $ 8.68 \times 10^{25}$~kg (\url{https://ssd .jpl .nasa.gov/}). The orbital elements used in the numerical simulations are presented in Table \ref{tab:1}. We also consider the two main gravitational terms of Titania, $J_2$ and $C_{22}$, with values equal to $1.13 \times 10^{-4}$ and $3,38 \times 10^{-5}$, respectively \cite{Chen2014}. They are the most critical terms in the gravitational field after the Keplerian term. These terms cause significant variations in some orbital elements, such as the eccentricity and inclination of the orbit. In this way, they become the most effective terms to model the irregular shape of a body \cite{Tzirt2009, Tzirt2010, Xavier2022}. In the case of Titania, it is worth noting that these terms are the only ones available in the literature. In addition to these coefficients, the mass of Titania ($35.27 \times 10^{20}$ kg) and its radius ($788.9$ km) are considered in this work. The equation of motion of the probe can be written in the following form \cite{Scheeres2006}:
\begin{equation}  
\begin{split}
 \ddot{\Vec{r}}= &- \frac{G(M_T+m)\Vec{r}}{\Vec{r}^3} + \\
& GM_U \left(\frac{\Vec{r}_U-\Vec{r}}{\mid \Vec{r}_U-\Vec{r} \mid ^3}-\frac{\Vec{r}_U}{\Vec{r}_U^3}\right) + \Vec{P}_T 
\end{split}
\label{eq:1}
\end{equation}

\begin{equation}
\begin{split}  
P_{Tx} = & - \frac{G(M_T+m) J_2x}{2r^5}\left[3-15\left(\frac{z}{r}\right)^2\right]\\
& + \frac{3 G(M_T+m) C_{22}x}{r^5} \left[2- \frac{5(x^2-y^2)}{r^2}\right] 
\end{split}
\label{eq:2}
\end{equation}
\begin{equation}
\begin{split}
P_{Ty} =  & - \dfrac{G(M_T+m) J_2y}{2r^5}\left[3-15\left(\dfrac{z}{r}\right)^2\right] \\ 
& - \dfrac{3 G(M_T+m) C_{22}y}{r^5} \left[2 + \dfrac{5(x^2-y^2)}{r^2}\right] 
\end{split}
\label{eq:3}
\end{equation}
\begin{equation}
\begin{split}
P_{Tz} =  & - \dfrac{G(M_T+m) J_2z}{2r^5}\left[9-15\left(\dfrac{z}{r}\right)^2\right] \\
& - \dfrac{15 G(M_T+m) C_{22}z}{r^7}(x^2-y^2) 
\end{split}
\label{eq:4}
\end{equation}

\noindent where $\vec{P}_T$ is the expansion of the gravitational potential up to the second order for Titania; $P_{Tx}$, $P_{Ty}$ and $P_{Tz}$ are components of the $P_T$ vector. $m$ is the mass of the space probe, $M_T$ the mass of Titania and $M_U$ the mass of Uranus. The terms $\vec{r}$ and $\vec{r}_U$ are the radius vector of the probe and Uranus, respectively. 

\begin{table}[ht]
\centering
  \caption{Parameters of Titania with respect to Uranus.}
 \label{tab:1}
 \begin{tabular}{cc}
 \hline
   \textbf{Parameter}         & \textbf{Value} \\
  \hline
  Semi-major axis (km)    &  $435.8 \times 10^{3}$     \\
  Eccentricity            &  $1.18 \times 10^{-3}$   \\  
  Inclination     ($^\circ$)   &  $1.0\times 10^{-1}$                    \\
  Argument of periapsis ($^\circ$)&  $1.64 \times 10^{2}$  \\
  Longitude of ascending node ($^\circ$) & $1.67 \times 10^{2}$    \\
  Mean anomaly  ($^\circ$)     & $2.05 \times 10^{2}$       \\
 
\hline
\end{tabular}\\
\hspace{-0.6cm}JPL.Website: \url{https://ssd.jpl.nasa.gov/}. 
 \end{table}

First, the orbits were {\bf numerically} integrated considering the system described by equations \ref{eq:1}-\ref{eq:4}. During these {\bf numerical } integrations, the maneuvers were performed. To integrate the orbits and to carry out the maneuvers, modifications were made to the Mercury package \cite{Chambers1999}.  


\subsection{Orbital Maneuvers}
\label{sub:manob}
During the decay process of an orbit, its orbital elements change. Therefore, when trying to perform a maneuver, we must consider some of these variables, such as the semi-major axis ($a$), eccentricity ($e$), and inclination ($I$). For example, suppose a maneuver is performed close to the probe's collision with the mission target's surface. In that case, the orbit will have a high eccentricity and, possibly, a considerable variation {\bf in its} semi-major axis. Thus, we present in this section two models of maneuvers. First, consider the transfer from an initial elliptical orbit to a final circular orbit and later, transferring the spacecraft from an initial elliptical orbit to a final elliptical orbit, thus correcting the original altitude of the orbit without considering the inclination. Both maneuvers are coplanar. The final orbit is the desired orbit for the mission, while the initial orbit is the orbit after the perturbations acted and changed the initial orbit. 

\subsubsection{Maneuvering from an elliptical to a circular orbit.}

The first maneuver presented here involves transferring the space probe from an elliptical to a final circular orbit. The first impulse ($\Delta V_1$) is applied at the apoapsis of the elliptical orbit, making the probe to enter a transfer ellipse. The second impulse ($\Delta V_2$) is applied when the probe is the apoapsis of the transfer orbit to make the probe occupy the desired final circular orbit. The total increment in velocity is given by the sum of the modules ($\Delta V_1$) and ($\Delta V_2$). The time for carrying out the orbital transfer is half the period of the transfer orbit. The equations for performing this maneuver are described as follows:
 \begin{equation}
 V_{ap}=\sqrt{\dfrac{2\mu}{a(1+e)}-\dfrac{\mu}{a}}
\end{equation}
    \begin{equation}
V_{p1}=\sqrt{\dfrac{2\mu}{a(1+e)}-\dfrac{2\mu}{a(1+e)+r_{cir}}}
    \label{eq:1.1} 
\end{equation}

\begin{equation}
 \Delta V_1=V_{p1}-V_{ap}
   \label{eq:2.1} 
\end{equation}

\begin{equation}
 V_{2}=\sqrt{\dfrac{2\mu}{r_{cir}}-\dfrac{2\mu}{a(1+e)+r_{cir}}}
\end{equation}
 
\begin{equation}
   V_{cir}=\sqrt{\dfrac{\mu}{r_{cir}}}
   \label{eq:4.1} 
\end{equation}

\begin{equation}
  \Delta V_2=V_{cir}-V_2
   \label{eq:5.1} 
\end{equation}

\begin{equation}
 \Delta V=\mid \Delta V_1 \mid + \mid \Delta V_2 \mid
   \label{eq:6.1} 
\end{equation}
\noindent where $V_{ap}$ is the velocity at the apoapsis of the initial orbit, $V_{p1}$ the velocity at point 1 in the transfer orbit, $\Delta V_1$ the first impulse, $V_2$ is the velocity at the apoapsis (point 2) of the transfer orbit, $V_{cir}$ the velocity in the final circular orbit, $\Delta V_2$ the second impulse and $\Delta V$ the total impulse. 

\subsubsection{Maneuvering between elliptical orbits}

Another maneuver presented in this work will be from an initial elliptical orbit to another final orbit, also elliptical. These orbits are coplanar. It happens when the nominal orbit of the spacecraft is elliptic. For this maneuver, the first impulse ($\Delta V_1$) is applied at the periapsis of the initial elliptical orbit. Then, the probe will occupy a transfer orbit. The second impulse ($\Delta V_2$) is applied at the apoapsis of the transfer orbit, and the spacecraft will be placed into its final elliptical orbit. The sum of ($\Delta V_1$) and ($\Delta V_2$) results in the total increment of velocity for performing the maneuver. The equations for carrying out such a maneuver are given by:
\begin{equation}
 V_{p1}=\sqrt{\dfrac{2 \mu}{r_{p1}}-\dfrac{2 \mu}{r_{p1}+r_{a1}}}
\end{equation}

\begin{equation}
 V_{pt}=\sqrt{\dfrac{2 \mu}{r_{p1}}-\dfrac{2 \mu}{r_{p1}+r_{ta}}}
\end{equation}

\begin{equation}
\Delta V_1 = V_{pt}-V_{p1}
\end{equation}

\begin{equation}
V_{at}=\sqrt{\dfrac{2 \mu}{r_{ta}}-\dfrac{2 \mu}{r_{p1}+r_{ta}}}
\end{equation}

\begin{equation}
V_{a2}=\sqrt{\dfrac{2 \mu}{r_{ta}}-\dfrac{2 \mu}{r_{p2}+r_{ta}}}
\end{equation}

\begin{equation}
\Delta V_2 = V_{a2}-V_{at}
\end{equation}

\begin{equation}
\Delta V = \mid \Delta V_1 \mid + \mid \Delta V_2 \mid
\end{equation}
\noindent where $r_{p1}$ and $r_{a1}$ are the initial orbit's periapsis and apoapsis. Furthermore, $r_{ta}$ is the apoapsis of the transfer ellipse, $r_{p2}$ is the periapsis of the final orbit, $r_{a2}=r_{ta}$ is the apoapsis of the desired final orbit. Regarding the increase in velocity, we have $V_{p1}$ equal to the velocity at point 1 of the initial orbit, $V_{pt}$ the velocity at point 1 of the transfer ellipse, $\Delta V_1$ the first impulse, $V_{at}$ the velocity at the apoapsis of the transfer ellipse, $V_{a2}$ the velocity at point 2 concerning the final desired orbit, $\Delta V_2$ the second impulse and $\Delta V$ the total impulse. 

\section{Analyzing the Results}
\label{sec:result}

\begin{figure}[H]
\centering
   \fbox{\includegraphics[width = 6.0cm]{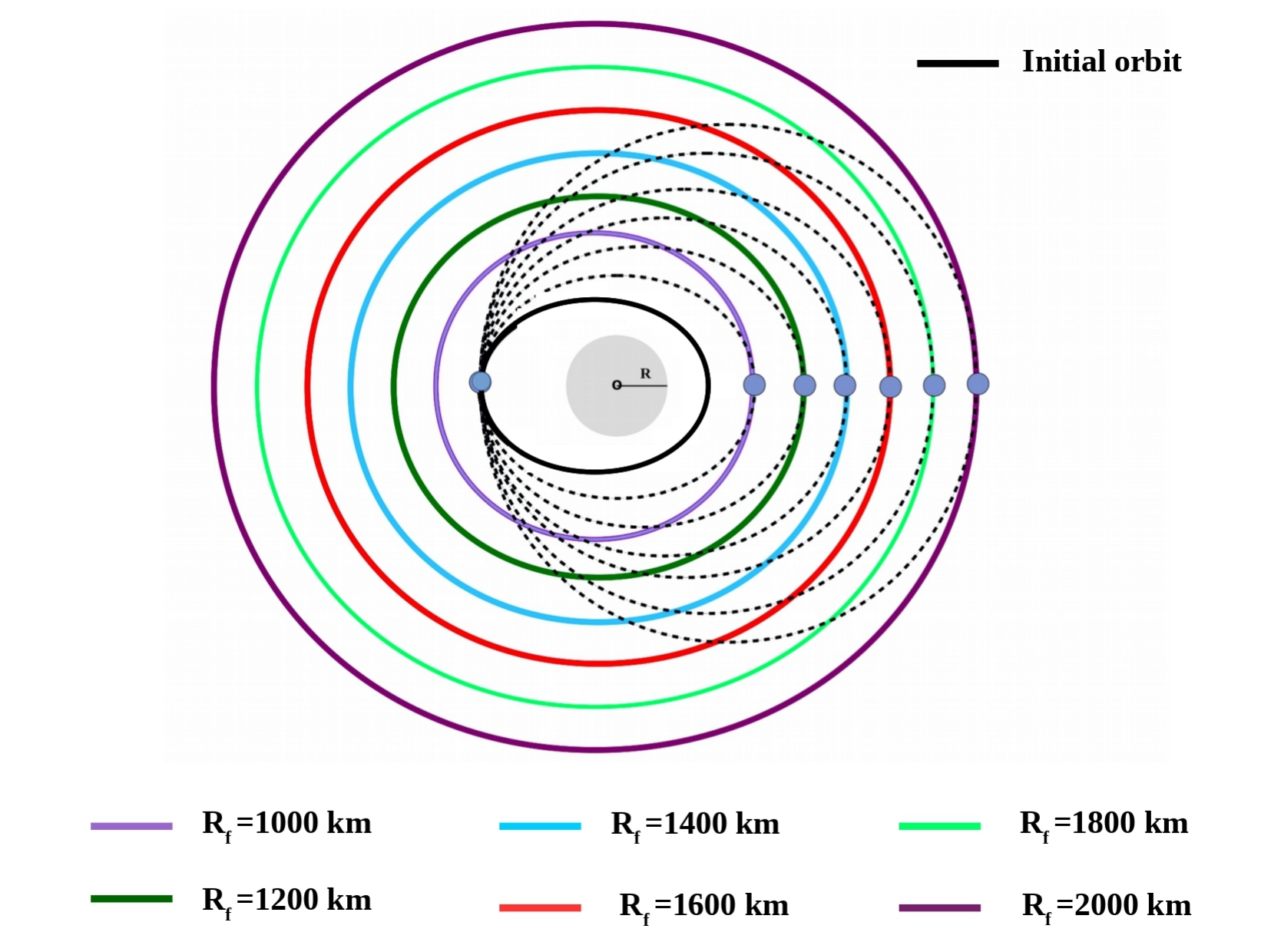}}
    \caption{Representation of the desired final circular orbits, where $R_f$ is the radius of these orbits. Purple for $R_f=1000$~km, green for $R_f= 1200$~km, blue for $R_f= 1400$~km, red for $R_f= 1600$~km, light green for $R_f= 1800$~km and dark purple for $R_f= 2000$~km. Initial orbit represented by the black line}.
    \label{fig:circular}
\end{figure}

The first results presented are about the first maneuver, from an initial elliptical orbit to a final circular orbit. This is the situation that occurs when the nominal orbit is circular and the orbit after the pertubations is elliptical. Then, the radii of the final desired orbits were chosen according to the regions with the most prolonged orbital duration, according to the analysis presented in \cite{Xavier2022}. This radius ranges from 1000-2000~km, in intervals of 200~km. The choice of the right moment to carry out the maneuver was made as follows: first, we chose a circular orbit with an initial semi-major axis, ($a_i=R_i$), ranging from 1000-2000 km, the we analyzed the decay of this orbit and, when reaching a significant value of variation from its starting position, the maneuver is performed. We take the orbital elements of this point and consider them as elements of the initial orbit (black orbit, Figure \ref{fig:circular}).
Small changes in an orbit's initial conditions can shorten its lifetime, and the orbit reaches high eccentricities and inclinations. Thus, some maneuvers are performed after a significant change in its initial position to return the orbit to its original altitude position, disregarding the inclination. Four initial inclinations were considered: $60^\circ$, $70^\circ$, $80^\circ$ and $90^\circ$. Figure \ref{fig:circular} shows a diagram of how the maneuvers will be carried out.  

To plan the maneuvers, we took into account the gravitational coefficients of Titania, including $J_2$ and $C_{22}$, as well as the gravitational coefficients of Uranus. After that, we executed the maneuvers necessary to achieve the desired orbits, utilizing equations \ref{eq:1.1}-\ref{eq:6.1}. 

The orbits with the largest deviations of 20-30\% of the initial value (black orbit, Figure \ref{fig:circular}), are those with initial inclinations of $60^\circ$, $70^\circ$ and $80^\circ$ for all simulations. In this case, when the orbits reach the deviation, the lifetime comes close to 240-250~days, and the $\Delta V$ needed to reach the circular orbit of radius 1000~km again is around $1 \times 10^{-2}-6 \times 10^{-2}$~km/s. Considering that orbits with this inclination have a total lifetime of 252~days, it is concluded that the maximum inclination variation occurs a few days before the collision.

In the case of an orbit with an initial inclination of $90^\circ$, the maximum variation of $I$ is 5\% for all final orbits. However, the $\Delta V$ ranges from $4 \times 10^{-2}-1 \times 10^{-1}$~km/s and the decay times range from 1-17~days. The smallest values of $\Delta V$ are for maneuvers to reach final orbits with a radius of 1000-1200~km. These orbits also have the highest rate of inclination variations.

The variations in the semi-major axis are quite small, between 0.1-0.9 \% of the initial value. For the four inclinations analyzed, $60^\circ$, $70^\circ$, $80^\circ$ and $90^\circ$, there are cases of maximum variations of the semi-major axis of 0.6 - 0.9 \%. In these cases, the $\Delta V$ ranges from $8 \times 10^{-2}-1.1 \times 10^{-1}$~km/s. The cases with the lowest value of $\Delta V$ are for final orbits with a 1000-1200 km radius.

\begin{figure}[h]
\centering
 \includegraphics[width=7.0cm]{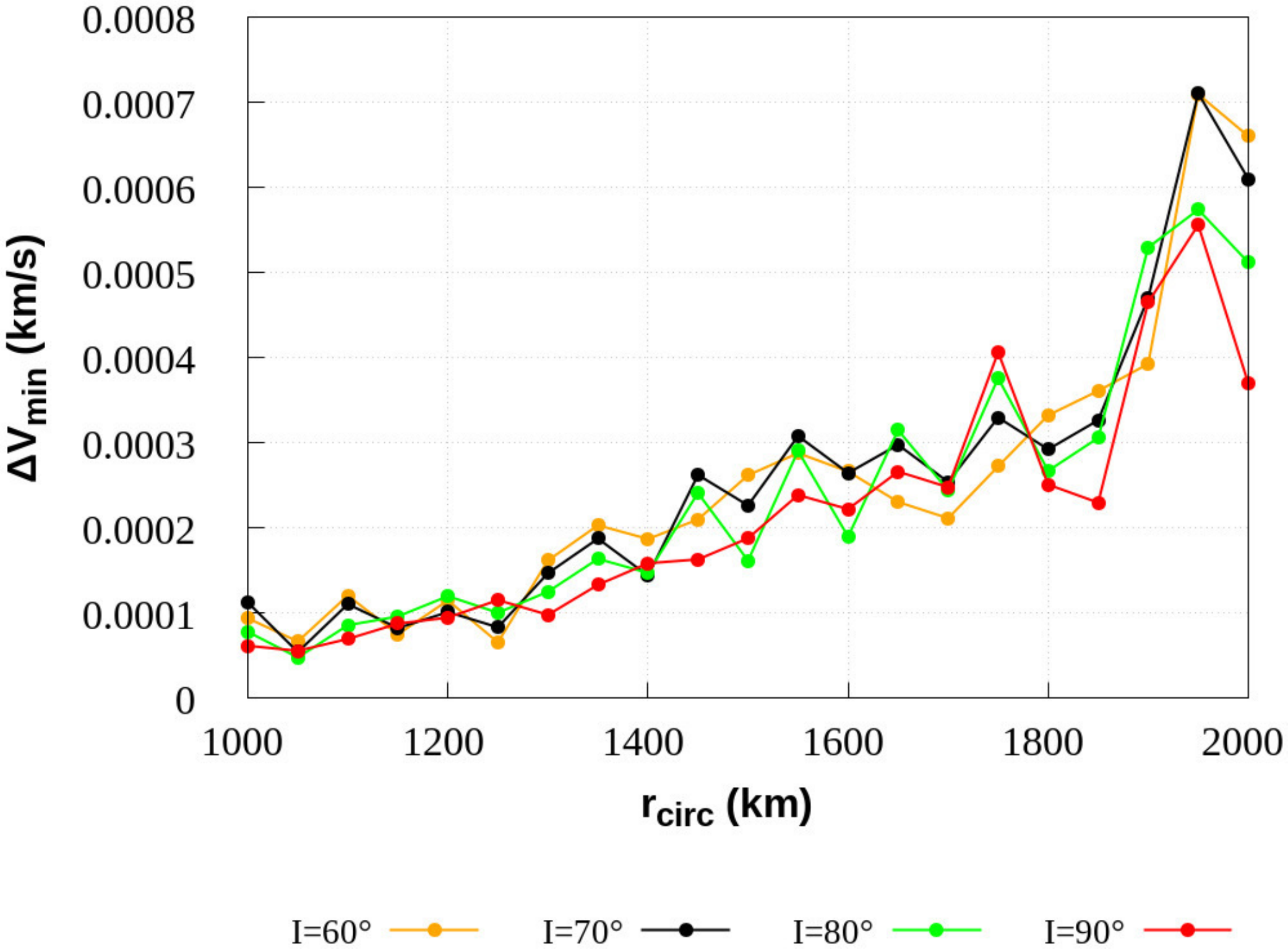}
\caption{Minimum $\Delta V$ as a function of the circular radius of the final orbit. Analyzing four initial inclinations $60^\circ$, $70^\circ$, $80^\circ$ and $90^\circ$ in orange, red, green, and black colors, respectively. Initial values: $a_i=R_i=1000-2000$~km, $e_i=0$ and $\omega= \Omega = 0^\circ$. The radius of intended final orbits is $1000-2000$~km.}
\label{fig:man_cir_3}
\end{figure}

Some missions choose to correct the spacecraft's position with a short lifetime before a considerable variation in its position. For example, Figure \ref{fig:man_cir_3} presents a study that seeks to find a minimum $\Delta V$ to perform an orbital transfer before a large variation of its initial position. Thus, now the maneuvers of interest are not the ones with a significant change in the inclination or in the semi-major axis. As in the previous cases, for the four inclinations analyzed, the minimum $\Delta V$ is calculated for cases where the final orbits have a 1000-2000 km radius. 
As shown in Figure \ref{fig:man_cir_3}, the smallest value of $\Delta V$  is approximately $5.4 \times 10^{-5}$~km/s, for a final orbit with a radius of 1050~km, for all inclinations. For the first inclination of $60^\circ$, the maneuver was performed after 5~days, equivalent to approximately 45 orbital periods. At that point, the semi-major axis and eccentricity had values equal to $1.0498 \times 10^{3}$~km and $2.8 \times 10^{-4}$, respectively. The ratio of change of the semi-major axis is $0.016$ \% of its initial value and $0.45$ \% for the inclination.

In the case of the orbit with an inclination of $70^\circ$, this maneuver was carried out after 14~days, when $a$ and $e$ are $1.0492 \times 10^{3} $~km and $2.3 \times 10^{-4}$, respectively. The inclination variation for this $\Delta V$ is $0.4$\%, and for the semi-major axis $0.01$ \%.
 
For $I$ equal to $80^\circ$ and $90^\circ$, the minimum $\Delta V$ is found after 5~days. In the case of $I=80^\circ$, at the moment of the maneuver, the value of the semi-major axis is $1.0497 \times 10^{3}$~km and the eccentricity $2.0 \times 10^{-4}$. The inclination and semi-major axis variations are $0.36$ \% and $0.02$ \%, respectively.
For an inclination of $90^\circ$, the maneuver is performed when $a=1.0497 \times 10^{3}$~km and $e=2.03 \times 10^{-4}$. The variation on the inclination is approximately $0.3$ \% and on the semi-major axis $0.023$ \%. Within this minimum range, it is possible to identify that; for a final orbit with a radius of 950~km, the $\Delta V$ required to perform the maneuver is greater than the other cases analyzed. For orbits with initial inclinations of $80^\circ$ and $90^\circ$, these values are close to $5.7 \times 10^{-4}$~km/s for $I=80^\circ$ and $5.55 \times 10^{-4}$~km/s for $I =90^\circ$. For $I=70^\circ$ this orbit can be achieved with  $\Delta V=7.1 \times 10^{-4}$~km/s.

\begin{table}
\centering
\caption{Values of $\Delta V$ for maneuvers with significant offset values in the inclination and in the semimajor axis. Initial conditions used in maneuvers such as initial inclination, orbital duration time, apoapsis radius ($r_{apo}$), final circular orbit radius ($r_{circ}$), semimajor axis offset (Def$_a$), and the inclination (Def$_I$) in percentage, and the equivalent $\Delta V$.}
\label{tab:dados_circ}
\scalebox{0.5}{
\begin{tabular}{@{}lllllll@{}}
\hline
$\mathbf{I (^\circ)}$  & \textbf{Time (Orbital period)}$^1$ & $\mathbf{r_{Apo}}$ (km) & $\mathbf{r_{circ}}$ (km) & \textbf{Def}$_a$ \textbf{(\%)} & \textbf{Def}$_I$ \textbf{(\%)} & $\mathbf{\Delta V}$ \textbf{(km/s)}\\ \hline

$I_i=60$ & 1735  & 1010  &   1000   &  0.057 & 24.9  & $2.5332 \times 10^{-3}$  \\

$I_i=70$ & 2172  & 1042  &   1000   &  0.44  & 28.3   & $1.0192 \times 10^{-2}$  \\ 

$I_i=80$ & 2264  & 1107 & 1000   & 0.07  & 25.4   & $2.5823 \times 10^{-2}$   \\ 

$I_i=90$ & 2236 & 1145  & 1000 &  0.06  & 2.0  & $3.5056 \times 10^{-2}$  \\ 
\hline

$I_i=60$ & 1735 & 1264  & 1200 & 0.067 & 14.7 & $1.1916 \times 10^{-2}$  \\
$I_i=70$ & 2136 & 1585  & 1200 & 0.06 & 20 & $7.0039 \times 10^{-2}$  \\ 

$I_i=80$ & 1872 & 1562 & 1200 & 0.13 & 13 & $6.6109 \times 10^{-2}$  \\ 

$I_i=90$ & 1800 & 1538 & 1200  & 0.04  & 1.8  & $6.1449 \times 10^{-2}$  \\ 
\hline
$I_i=60$ & 1934 & 1990  & 1400 & 0.084 & 14.5  & $8.5579 \times 10^{-2}$  \\ 
$I_i=70$  & 1581 & 1990  & 1400 & 0.11 &  13.3 & $8.5653 \times 10^{-2}$  \\ 

$I_i=80$ & 1436 & 1948 & 1400 & 0.11 & 7.6 & $7.9481 \times 10^{-2}$  \\ 

$I_i=90$ & 1400 & 1912 & 1400 & 0.09 & 1.9 & $7.4107 \times 10^{-2}$  \\ \hline 
$I_i=60$ & 1462 & 2348 & 1600 & 0.072 & 14.6 & $8.8944 \times 10^{-2}$  \\
$I_i=70$ & 1227 & 2383 & 1600 & 0.19 & 11 & $9.3866 \times 10^{-2}$  \\ 
$I_i=80$ & 1118 & 2302 & 1600 & 0.18 & 6.0 & $8.3743 \times 10^{-2}$  \\ 

$I_i=90$  & 1127 & 2332 & 1600 & 0.3  &  2.2  & $8.7669 \times 10^{-2}$  \\ 
\hline
$I_i=60$ & 1144 & 2700  & 1800 & 0.22 & 14.6  & $9.0549 \times 10^{-2}$  \\ 

$I_i=70$ & 963 & 2653  & 1800 & 0.5 &  8.6  & $8.6478 \times 10^{-2}$  \\ 

$I_i=80$ & 918 & 2784 & 1800 & 0.3  & 6.8   & $9.9839 \times 10^{-2}$  \\ 

$I_i=90$ & 890 & 2626 & 1800 & 0.3 & 2.4 & $8.3109 \times 10^{-2}$  \\ 
\hline
$I_i=60$ & 953  & 3094 & 2000 & 0.76 & 17.4 & $9.6332 \times 10^{-2}$  \\ 
$I_i=70$ & 800  & 3125 & 2000 & 0.4  & 12  & $9.8090 \times 10^{-2}$  \\ 
$I_i=80$ & 727  & 2990 & 2000 & 0.4 & 6.0 & $8.5558 \times 10^{-2}$  \\ 
$I_i=90$ & 727 & 3006 & 2000 & 0.4  & 3.2 & $8.7159 \times 10^{-2}$  \\ \hline
\end{tabular}}\\
\footnotetext[1]{The orbital period is measured from the period of the orbit closest to the surface of the central body.}
\end{table}

In order to analyze the best values of $\Delta V$ used to perform orbital maneuvers, Table \ref{tab:dados_circ} shows some values of $\Delta V$ used for orbits lagged in inclination and semimajor axis. For each inclination, $I=60^\circ$, $I=70^\circ$, $I=80^\circ$ and $I=90^\circ$, values of $\Delta V$ are presented, as well as the initial conditions for carrying out the maneuvers. Cases are presented for final orbits with radii equal to 1000, 1200, 1400, 1600, 1800, and 2000 km.

\begin{figure}[H]
\centering
\fbox{\includegraphics[width=5.0cm]{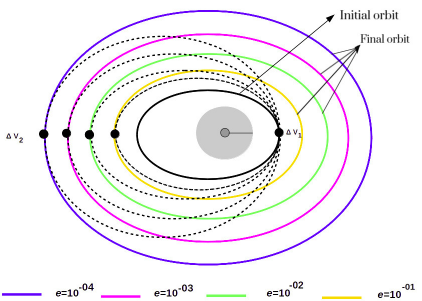}}
\caption{Representation of final elliptical orbits. Orange for eccentricity $10^{-1}$, green for  $10^{-2}$, pink for $10^{-3}$ and blue for  $10^{-4}$.}
\label{fig:man_el_2}
\end{figure}

The next discussions are about the results obtained for maneuvers between two elliptical orbits. In \cite{Xavier2022}, the orbital decay of objects with eccentricity equal to $10^{-4}$, $10^{-3}$, $10^{-2}$ and $10^{-1}$ are studied. The authors concluded that the lifetime was longer for orbits with an eccentricity $10^{-3}$. Thus, the intended final orbits investigated in this work have the same eccentricity as the orbits studied in that work ($10^{-4}$, $10^{-3}$, $10^{-2}$ and $10^{ -1}$). We call these eccentricities $e_f$. Figure \ref{fig:man_el_2} shows these orbits.
A set of numerical simulations was made to perform the maneuvers for each eccentricity.

Six points in the orbit were chosen to carry out the maneuvers for each value of $e$. This choice aims to analyze whether it is more feasible to make one maneuver at different times, resulting in several maneuvers before the collision, or only one maneuver moments before the spacecraft collides with the surface of the moon. Such maneuvers are called $\mathbf{m_P}$, $\mathbf{m_1}$, $\mathbf{m_2}$, $\mathbf{m_3}$, $\mathbf{m_4}$  and  $\mathbf{m_5}$. Being $m_P$, the main maneuver is because this maneuver is performed at the closest point to the surface of Titania, that is, closest to the collision. The approximate periapsis for this maneuver is 790~km for all values of $e$. The point for maneuver $m_5$ is farthest from the surface of Titania, with the maneuver performed with a few days of decay compared to $m_P$. The periapsis for $m_5$ is approximately 965~km. The $m_1$, $m_2$, $m_3$, and $m_4$ maneuvers are carried out at intermediate points between $m_P$ and $m_5$. The periapsis of $m_1$ is around 825~km, $m_2 \approx 860$~km, $m_3 \approx 895$~km, $m_4 \approx 930$~km and $m_5 \approx 965$~km. Points vary every 35~km.

All simulations were made considering $a=1000$~km, $I=90^\circ$, $\omega=\Omega=0^\circ$, the gravitational coefficients  $J_2$ and $C_{22}$ of Titania and the gravitational effects of Uranus.

\begin{figure}
\centering
    \fbox{\includegraphics[width=7.0cm, height = 5.40cm]{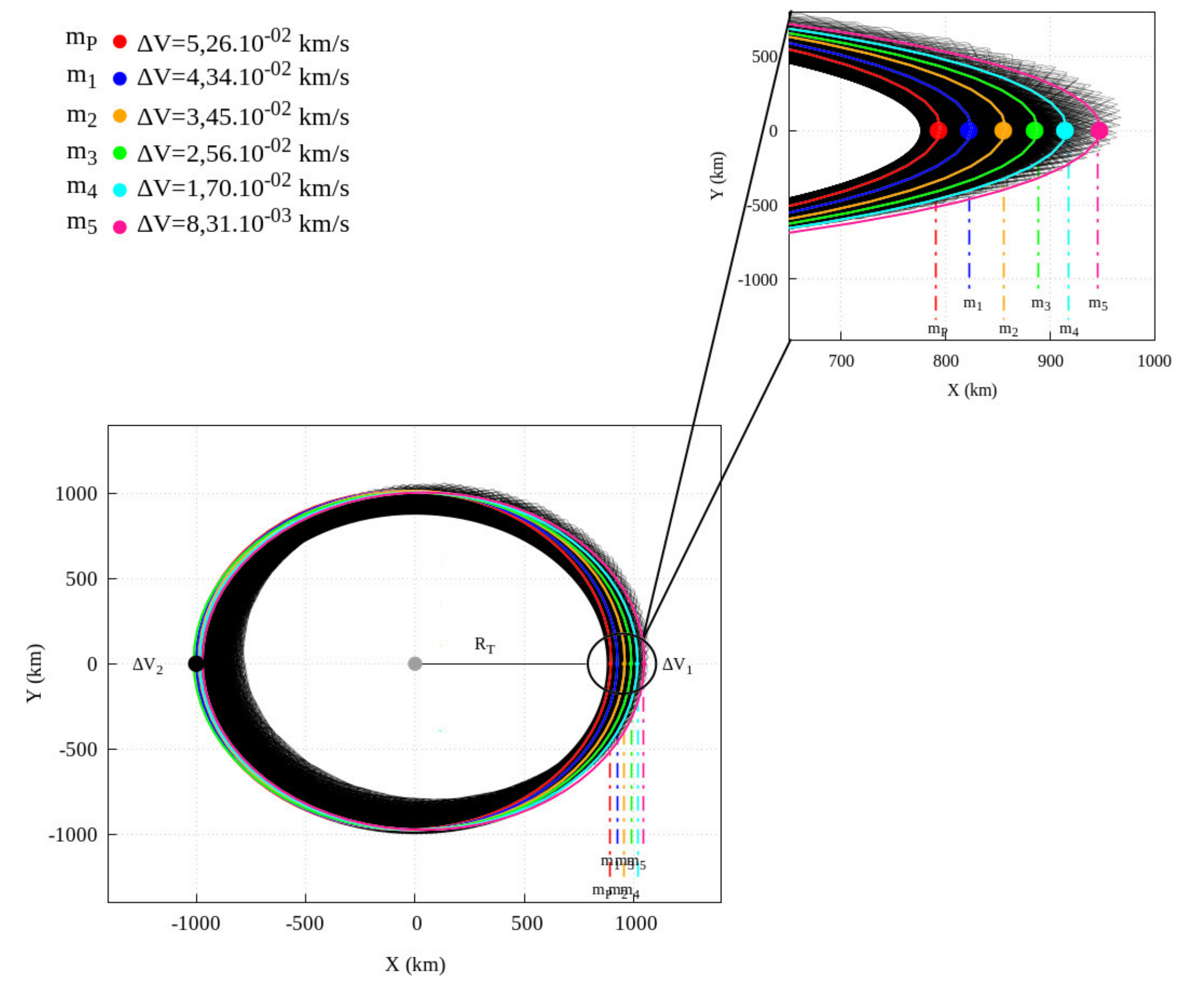}} \fbox{\includegraphics[width=7.0cm]{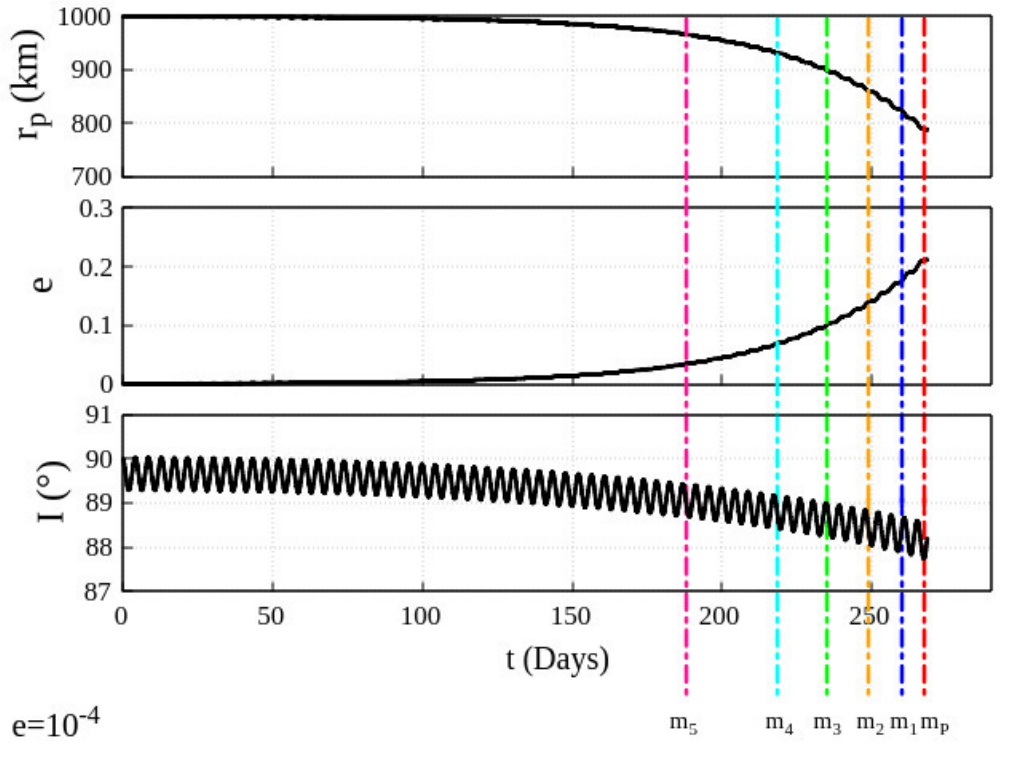}}

 \caption{Projection of orbital positions for the six maneuvers performed ($m_P$, $m_1$, $m_2$, $m_3$, $m_4$ and $m_5$) for $e~=~10^{-4}$ (at the top). Variation of orbital elements $e$, $I$, and periapsis radius as a function of lifetime and maneuvering time (at the bottom). Initial conditions: $a=1000$~km, $e=10^{-4}$, $I=90^\circ$, $\omega=\Omega= 0^\circ$.}
 \label{fig:man_el_3}
\end{figure}

The first set of maneuvers was done for $e~=~10^{-4}$, giving a total lifetime of 268.536~days. All points of the orbit considered for  the maneuvers are represented in Figure \ref{fig:man_el_3} (at the top). To relate the variations of some parameters when performing the maneuvers, Figure \ref{fig:man_el_3} (at the bottom) shows the position of the periapsis  ($r_p$), $e$ and $I$ as a function of time. The day each maneuver was performed can also be visualized in this figure.

For the main maneuver $m_P$, the parameters are: $a=9.993 \times 10^{2}$~km, $e=2.094 \times 10^{-1}$, $I=87.9^\circ $, periapsis equal to 790.0~km, apoapsis equal to 1208.59~km and maneuver time of 267.26~days. Considering these values, we decay approximately 21\% in the periapsis, 2.3\% in the inclination, and 0.1 \% in the semi-major axis. There is also an increase in the eccentricity of three orders. The $m_P$ maneuver was performed about a day before the collision. The estimated $\Delta V$ for this transfer was $\approx 5.26 \times 10^{-2}$~km/s. The representation of the location of this maneuver can be seen in Figure \ref{fig:man_el_3}, in a dotted red line. 

The second maneuver ($m_1$), for $e=10^{-4}$, was performed within 259.87~days of numerical  integration, approximately 8~days before the collision. At this chosen point, the periapsis of the orbit was 825.09~ km, approximately 17.5\% of its initial value. The other initial conditions were: $a=9.989 \times 10^{2}$~km, $e=1.740 \times 10^{-1}$, $I=88.3^\circ$, and apoapsis equal to 1172.88~km. To perform this maneuver, the $\Delta V$ used was approximately $4.34 \times 10^{-2}$~km/s. Analyzing the variation of the orbital elements for this maneuver, it is possible to observe that, during 259.87~days, the periapsis decreased by almost 180~km, the inclination by $1.4^\circ$ and the semi-major axis by 2~km. The maneuver $m_1$ is represented by the dotted blue line in Figure \ref{fig:man_el_3}.

The maneuver ($m_2$) was performed after 248.59~days, about 19~days before the collision. In this maneuver, the $\Delta V$ required to reach the orbit with an eccentricity of $10^{-4}$ was $3.45 \times 10^{-2}$~km/s. The following initial conditions were used to execute the maneuver: $a=9.991 \times 10^{2}$~km, $e=1.392 \times 10^{-1}$, $I=88.5^\circ$, periapsis equal to 860.01~km, and apoapsis equal to 1138.35~km. The periapsis had the highest percentage of change of all the variations of the orbital elements during the decay of the orbit. It decayed almost 14\%. The semi-major axis and inclination decreased by 0.1\% and 1.66\%, respectively. In the case of eccentricity, there was an increase of three orders of magnitude about its initial value. The line orange represents this maneuver in Figure \ref{fig:man_el_3}. 

In figure \ref{fig:man_el_3} the color representing the maneuver ($m_3$) for $e=10^{-4}$ is green. The lifetime for this maneuver was 234.86 days, almost 33 days before the collision. In this maneuver, the periapsis decay was approximately 10\%, from an initial value of 999.9~km to 895~km. The eccentricity increased by two orders of magnitude, to $e=9.970 \times 10^{-2}$, and the apoapsis increased by almost 103~km. The semi-major axis and inclination had small variations, 0.1\% and 1.22\%, respectively. The maneuver $m_3$ has a smaller offset of the orbital elements concerning the previous maneuvers. However, the point where the maneuver is carried out may be of interest to a mission, and the $\Delta V$ necessary is $2.56 \times 10^{-2}$~km/s.  

In the case of the maneuver ($m_4$), represented in Figure \ref{fig:man_el_3} by the light blue color, the time to perform the maneuver was 218.20~days. The maneuver preceded 50.34~days of the collision, and its periapsis was 930.16~km, almost a drop of 7\% compared to its initial value. Regarding the inclination, the variation was small, around 1.33\%. The eccentricity with a value of $6.95 \times 10^{-2}$ and a $\Delta V$ of $1.70 \times 10^{-2}$~km/s was used to reach the same origin position.

In pink, the last maneuver for $e=10^{-4}$ shown in Figure \ref{fig:man_el_3} was performed when the orbit reached $187.84$~days. As already mentioned, this maneuver is performed within fewer days since the start of the mission than the others and, therefore, has a smaller periapsis decay. The initial conditions for this point are $a=9.991\times10^{2}$~km, $e=3.420 \times 10^{-2}$, $I=89^\circ$, periapsis equal to 965.01~km and apoapsis equal to 1033.37~km. The periapsis offset was $\approx 3.5\%$ and the inclination $\approx 1.11\%$, with a velocity increment $\Delta V= 8.31 \times 10^{-3}$~km/s. 

Comparing all maneuvers performed for $e=10^{-4}$, it is possible to observe that the best maneuver is ($m_5$). The $\Delta V$ of $8.31 \times 10^{-3}$~km/s is much smaller for the number of days of life. Compared to the other maneuvers, ($m_P$) for example, if two maneuvers ($m_5$) are performed, the orbit can have a lifetime of up to 374~days with 2$\times\Delta V$ equal to $1.662 \times 10^{-2}$~km/s. While two maneuvers ($m_P$), we will have a lifetime of $\approx 534$~days but a total $\Delta V$ of $1.052 \times 10^{-1}$~km/s, almost 6 times higher than the total $\Delta V$ used in ($m_5$). If we do six maneuvers at the point of the maneuver ($m_5$), we would have a longer lifetime and lower fuel consumption. This comparison holds for the other maneuvers as well. 

The following discussions are for calculating $\Delta V$ for $e=10^{-3}$. Since the results are closer to the case presented for $e=10^{-4}$, we will omit the figure with the projections of the maneuvering points. As in the previous case, six points in the orbit were chosen to carry out the maneuver. In this case, the orbital duration time was 313~days.

The $m_1$ maneuver took place with a duration of 304.44~days, 8.56~days before the collision. For this maneuver, a periapsis of 825.18~km was considered, 17.4\% smaller than its initial value, and an apoapsis of 1157~km, almost 15\% larger than the value before the decay. The other initial conditions were: $a=9.999\times10^{2}$~km, $e=1.747\times10^{-1}$, $I=87.9^\circ$. The value of $\Delta V$ was approximately $4.33\times10^{-2}$~km/s. Comparing the value of $\Delta V$ found for $m_1$ in the previous case, it is noticed that the values are very close, and the differences are irrelevant.

For maneuver $m_2$, we used the following initial conditions: maneuver time of 295.15~days, almost 18~days before the collision. $a=9.996\times10^{2}$~km, $e=1.396\times10^{-1}$, $I=88.0^\circ$, periapsis  equal to 860.10~km and apoapsis  equal to 1139.22~km. Regarding the variations of the parameters, we had a decay of almost 14\% in the periapsis and 2.2\% for the inclination. With these initial conditions, we found a $\Delta V$ of $\approx 3.43\times10^{-2}$~km/s. As in the previous maneuvers, the value of $\Delta V$ for $m_2$ is very close to that of $\Delta V$ found for the same maneuver analyzed in the case $e=10^{-4}$.  

In $m_3$, the $\Delta V$ found was $2.54\times10^{-2}$~km/s, in a time of 282.04~days, which preceded 30.96~days of the collision. At this point, the orbit has a semi-major axis of $9.990\times10^{2}$~km, eccentricity $1.041\times10^{-1}$, and inclination equal to $88.2^\circ$, with a periapsis variation of 10.4\% and apoapsis with an increase of almost 10.2\%. The maneuver $m_3$ also has a similar value to the one analyzed for $e=10^{-4}$. For $m_4$, the value of $\Delta V$ found was $\approx 1.68\times10^{-2}$~km/s for a lifetime of 264.93~days. In these maneuvers, the periapsis had a decay of almost 7\% and an inclination of $\approx 1.5\%$. Periapsis is equal to 930.03~km, apoapsis is equal to 1069.74~km, and eccentricity is equal to $6.986\times10^{-2}$. For the same maneuver analyzed in $10^{-4}$, we can say that the values of $\Delta V$ are closer.

 The initial conditions for the orbital transfer for $m_5$ were considered: $a=9.989\times10^{2}$~km, $e=3.388\times10^{-2}$, $I=88.9^\circ$, periapsis at 965.05˜km and apoapsis at 1032.76~km. The results showed that $\Delta V=8.01\times10^{-3}$~km/s, a decay of $\approx 3.4\%$ from the periapsis and $\approx 1.22\%$ for inclination and a lifetime of 234.19 days, almost 79 days before the collision. 

Carrying out a similar analysis to the one made for $e=10^{-4}$, this time between $m_5$ and $m_1$, we note that the maneuver $m_5$ is the most viable of all. While $m_5$ spent $8.01\times10^{-3}$~km/s to perform a maneuver after 234~days of life, $m_1$ spent $4.33\times10^{-2}$~km/ s for 304.44~days of life. If we do two maneuvers $m_5$, we will have the probe in orbit for 468.38~days spending a total $\Delta V$ of $1.602\times10^{-2}$~km/s. In the case of two maneuvers $m_1$, the total $\Delta V$ would be $8.66\times10^{-2}$~km/s for a little more than 608~days of life, almost 5.4 times greater than the $ \Delta V$ used for two maneuvers $m_5$.

Graphically, the results shown for $e=10^{-2}$ are similar to those for $e=10^{-1}$, which we will show below, and, for this reason, we omitted it here. For $e=10^{-2}$, the maneuver $m_P$ was performed with a $\Delta V$ of $4.99\times10^{-2}$~km/s. Compared to the values of $\Delta V$ for $m_P$  in the cases for $e$ equal to $10^{-4}$ and $10^{-3}$, this value is slightly lower, around $5\%$. In this maneuver, the periapsis offset was 20.2\%, and only 0.6\% for the inclination. The other initial conditions used were: $a=9.988\times10^{2}$~km, $e=2.089\times10^{-1}$, $I=89.4^\circ$, apoapsis  equal to 1207.00~km. The maneuver occurred at 152.68~days, 0.5~days before the collision with the central body.

In the case of $m_1$, the maneuver time preceded almost seven days of the collision (145.90~days), with a decay of almost 17\% at the periapsis  (825.01~km) and 1.2\% in the inclination. The $\Delta V$ needed to perform the transfer was $\approx 4.10\times10^{- 2}$~km/s. The semi-major axis was $9.995\times10^{2}$~km, eccentricity equal to $1.745\times10^{-1}$ and apoapsis  equal to 1174.00~km. The maneuver $m_1$ also has a slightly lower value than the maneuvers analyzed in the previous cases, around 5\%, for the same periapsis.

The turnaround time for $m_2$ was 135.00~days, 18~days before the end of the numerical integration. To perform this maneuver, a $\Delta V$ of $\approx 2.92\times10^{-2}$~km/s was found. Comparing with the $\Delta V$  for the cases $e=10^{-4}$ and $e=10^{-3}$, $\Delta V$ is now almost 15\% smaller. The initial elliptical orbit has a $a=9.991\times10^{2}$~km, $e=1.389\times10^{-1}$, $I=89.6^\circ$, periapsis equal to 860.26~km ( $\approx 13\%$ less than the initial value) and apoapsis equal to 1137.90~km ($\approx 13\%$ greater than the initial value).

For the maneuver $m_3$ performed approximately 30~days before the collision, the orbit had the following initial conditions: $a=9.989\times10^{2}$~km, $e=1.039\times10^{-1}$, $I =89.5^\circ$, periapsis equal to 895.07~km, with a decay of almost 9\% and an apoapsis of 1102.80~km, almost 93~km greater than its initial value. The total impulse in velocity was $\approx 2.31\times10^{-2}$~km/s. This value is almost 10\% less than the values of $\Delta V$ found in cases $e=10^ {-4}$ and $e=10^{-3}$. 

The maneuver represented in light blue, $m_4$, happened in 104.85~days. To find the increase in velocity $\Delta V$ at this point, a periapsis with a variation of 6\% was considered. The inclination had a small variation, about $0.3\%$, until the moment of the maneuver. With these initial conditions,  semi-major axis equal to $9.993\times10^{2}$~km,  $e = 6.931\times10^{-2}$, $I=89.7^\circ$, periapsis  equal to 930.07~km and apoapsis equal to 1068.6~km, a $\Delta V$ of $1.45\times10^{-2}$~km/s was found. This value is also considerably smaller than the ones found for $m_4$ in the previous cases.

The last maneuver, $m_5$, for $e=10^{-2}$, was performed after 74.11~days of life. The initial conditions were: $a=9.999\times10^{2}$~km, $e=3.482\times10^{-2}$, $I=89.8^\circ$, periapsis  equal to 965.14~km with decay of 2.5\% and apoapsis  equal to  1034.7~km. In this case, the $\Delta V$ was $6.04\times10^{-3}$~km/s.

In all maneuvers performed for a final orbit with eccentricity $10^{-2}$, the $ \Delta V$ presented lower values than those found for the same maneuvers with $e=10^{-4}$ and $e=10^{-3}$. In terms of percentage, we can say that $m_P \approx 5\%$, $m_1 \approx 5\%$, $m_2 \approx 15\%$, $m_3 \approx 9\%$, $m_4$ between 15\% and 17\% and $m_5$ between 25\% and 27\% smaller for $e=10^{-2}$ compared to the previous cases.

Maneuver $m_5$ appears, again, as the best maneuver found. Here, we will compare this maneuver with $m_2$. Doing two maneuvers $m_5$ will give an orbit with a lifetime of approximately 148~days, with a total $\Delta V$ of $1.208\times10^{-2}$~km/s. However, doing two maneuvers $m_2$, the orbit will have a lifetime of 270~days using a total $\Delta V$ of $6.38\times10^{-2}$~km/s. Although the lifetime is longer for $m_2$, the $\Delta V$ for $m_5$ is almost 5.2 times smaller.
\begin{figure}
\centering
 \fbox{\includegraphics[width=7.0cm, height = 5.5cm]{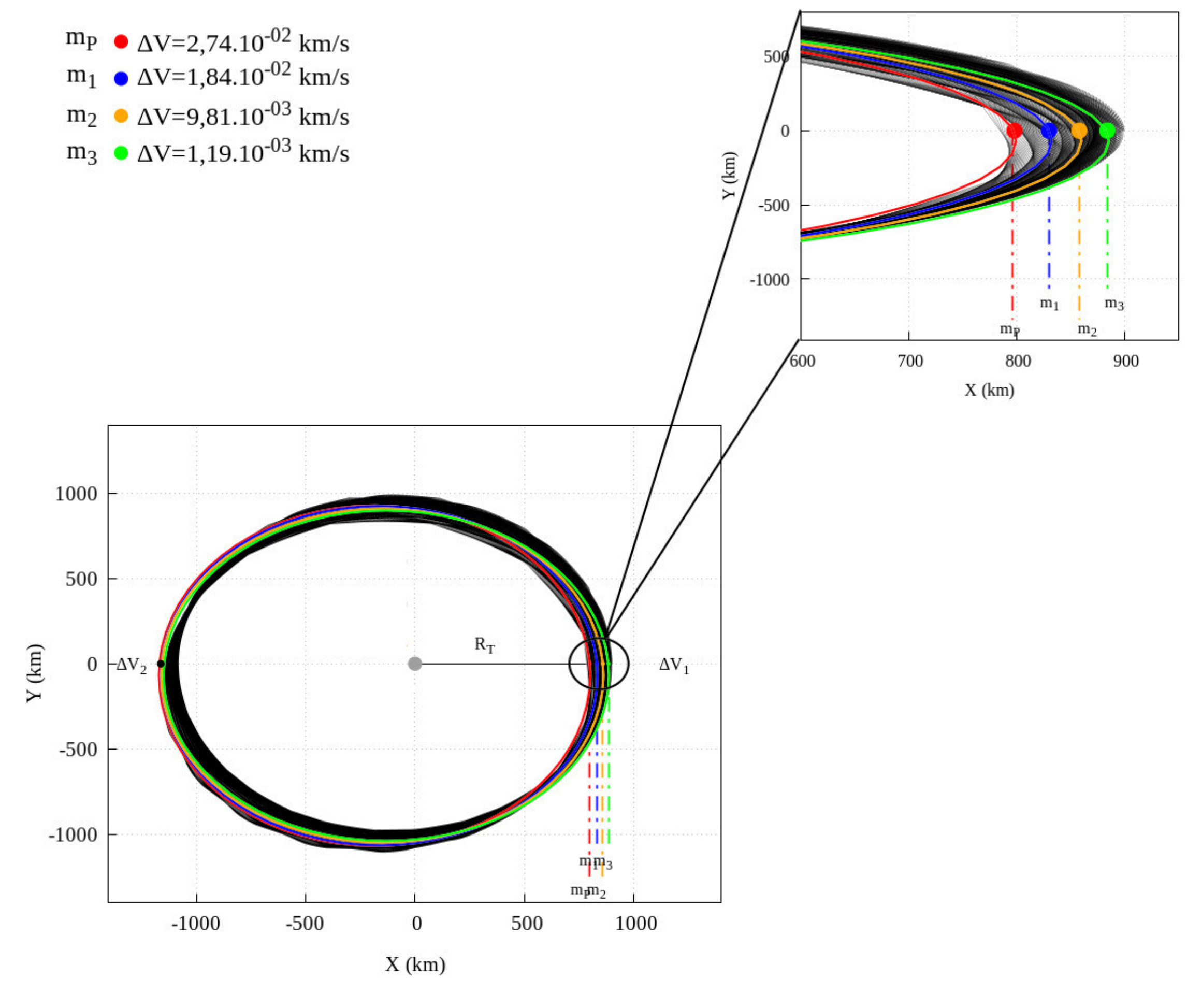}} \fbox{\includegraphics[width=7.0cm]{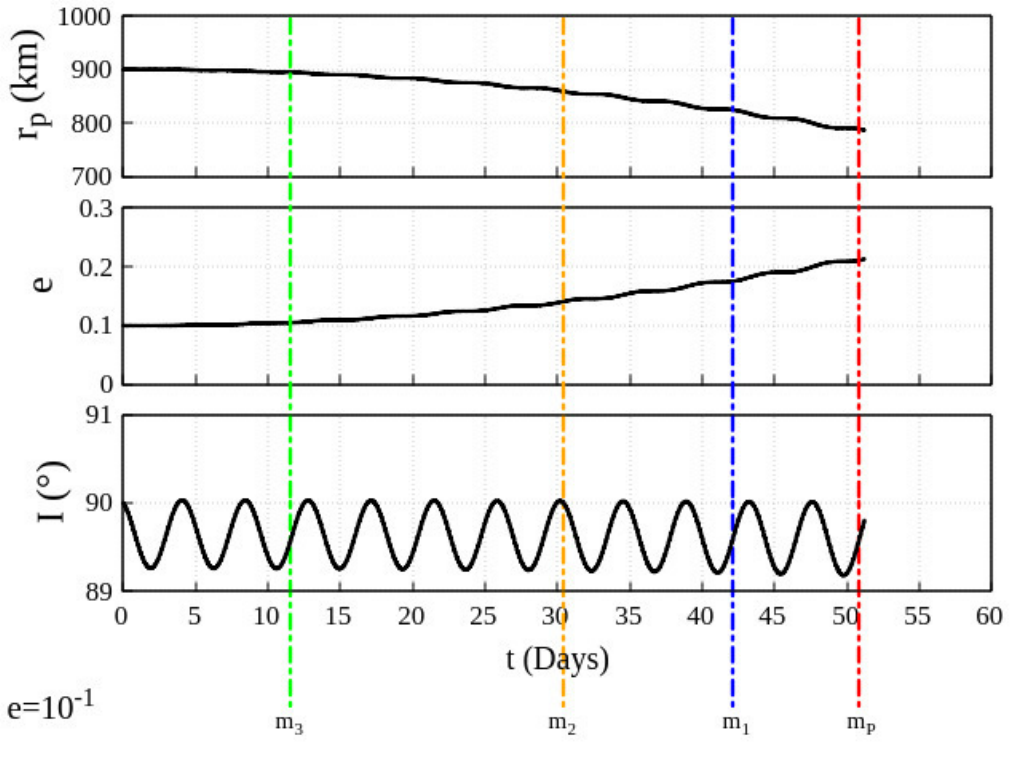}}

\caption{Projection of orbital positions for four maneuvers performed ($m_P$, $m_1$, $m_2$ and $m_3$) for $e=10^{-1}$ (at the top). Variation of orbital elements $e$ and $I$ and periapsis radius as a function of lifetime and maneuvering time (at the bottom). Initial conditions: $a=1000$~km, $e=10^{-1}$, $I=90^\circ$, $\omega=\Omega= 0^\circ$. Simulation time equal to 52~days.}
\label{fig:man_el_6}
\end{figure}

Finally, we present in Figure \ref{fig:man_el_6} the results for the maneuvers performed for $e=10^{-1}$. For this case, we consider only four maneuvers, $m_P$, $m_1$, $m_2$, and $m_3$. This choice is justified because the initial periapsis of the orbits starts at 990~km, so the points for maneuvers $m_4$ (periapsis equal to 930~km) and $m_5$ (periapsis equal to 965~km) are not possible. The colors and points are the same considered in the previous cases for $m_P$ to $m_3$. The total integration time, in this case, was 52 days.

The $m_P$ maneuver was performed considering a periapsis of 790~km and an apoapsis =1209.5~km. We adopted semi-major equal to $9.998\times10^{2}$~km, $e=2.098\times10^{-1}$ and $I=89.5^\circ$. The total impulse of $2.74\times10^{-2}$~km/s was given almost a day before the collision. At the time of the maneuver, the periapsis offset was approximately 12.2\% and 0 .55\% for the inclination. Compared to the values of $\Delta V$ analysed for $m_P$ in the previous cases, the value found is 45\% smaller than the value found for $e=10^{-2}$, almost 48\% smaller for $e =10^{-3}$ and for $e=10^{-4}$.  

In the case of $m_1$, the increase in velocity was given almost eight days before the collision, in 42.12~days. The value of $\Delta V=1.84\times10^{-2}$~km/s is approximately 55\% smaller than the value found for the case $e=10^{-2}$, 57\% smaller for $e=10^{-3}$ and 57.6\% smaller for $e=10^{-4}$. For this transfer, the  value $a$ of the orbit was $9.994\times10^{2}$~km, $e=1.744\times10^{-1}$, $I=89.5^\circ$, periapsis equal to 825.07~km, apoapsis equal to 1173.7~km and maneuver time 42.12~days.

The time to perform the maneuver $m_2$ was 30.46 days, a little more than 21 days before the collision. The value of $\Delta V$ used to perform the orbit transfer was $\approx 9.81\times10^{-3}$~km/s. At the time of the maneuver, the initial orbit had a periapsis of 860.0~km, 4.44\% variation about its initial value. The value of $a$ equal to $1.0\times10^{3}$~km, $e=1.399\times10^{-1}$, $I=89.9^\circ$ and apoapsis of 1140.0~km, almost 4\% higher than its initial value. The $\Delta V$ used in this maneuver is significantly smaller than those found for $m_2$ in the previous cases. Compared to the maneuvers $m_2$ performed for the previous cases, this maneuver had a $\Delta V$ 71.5\% and 71.4\% lower compared to the cases $e=10^{-4}$ and $e =10^{-3}$, respectively. For $e=10^{-2}$ this difference was 66.4\% smaller.

The last maneuver $(m_3)$ for $e=10^{-1}$ had the smallest value of $\Delta V$ found in relation to all analyzed cases, $\approx 1.19\times10^{-3 }$~km/s. The high value of the eccentricity makes the collision with the central body happen within a few days of integration, so the maneuver is performed within a few days of life. In this case, the transfer occurred with only 11.53 days of orbital duration. To perform this maneuver, the following initial conditions were considered, $a=9.998\times 10^{2}$~km, $e=1.048\times10^{-1}$, $I=89.5^\circ$, apoapsis equal to 895.0~km and an apoapsis of 1104.0~km. As mentioned, this maneuver had the lowest cost to transfer from one elliptical orbit to another. Compared to the same maneuver performed for $e=10^{-4}$, this maneuver has a $\Delta V$ 95\% smaller for $e=10^{-3}$ this percentage is 95.3\% and, 94.8\% for $e=10^{-2}$.
\begin{figure}
\centering
   \fbox{\includegraphics[width=6.0cm]{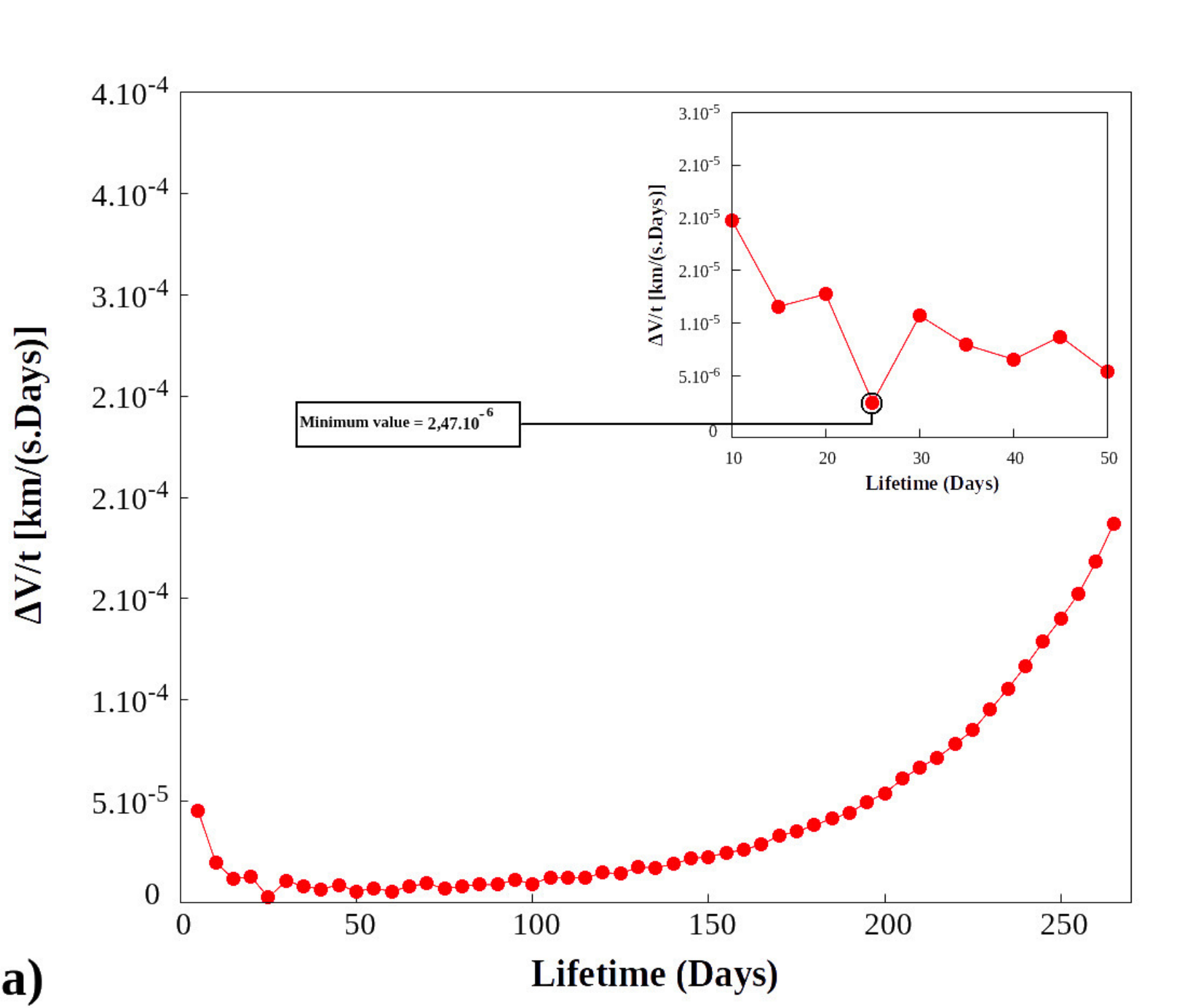}}
   \fbox{\includegraphics[width=6.0cm]{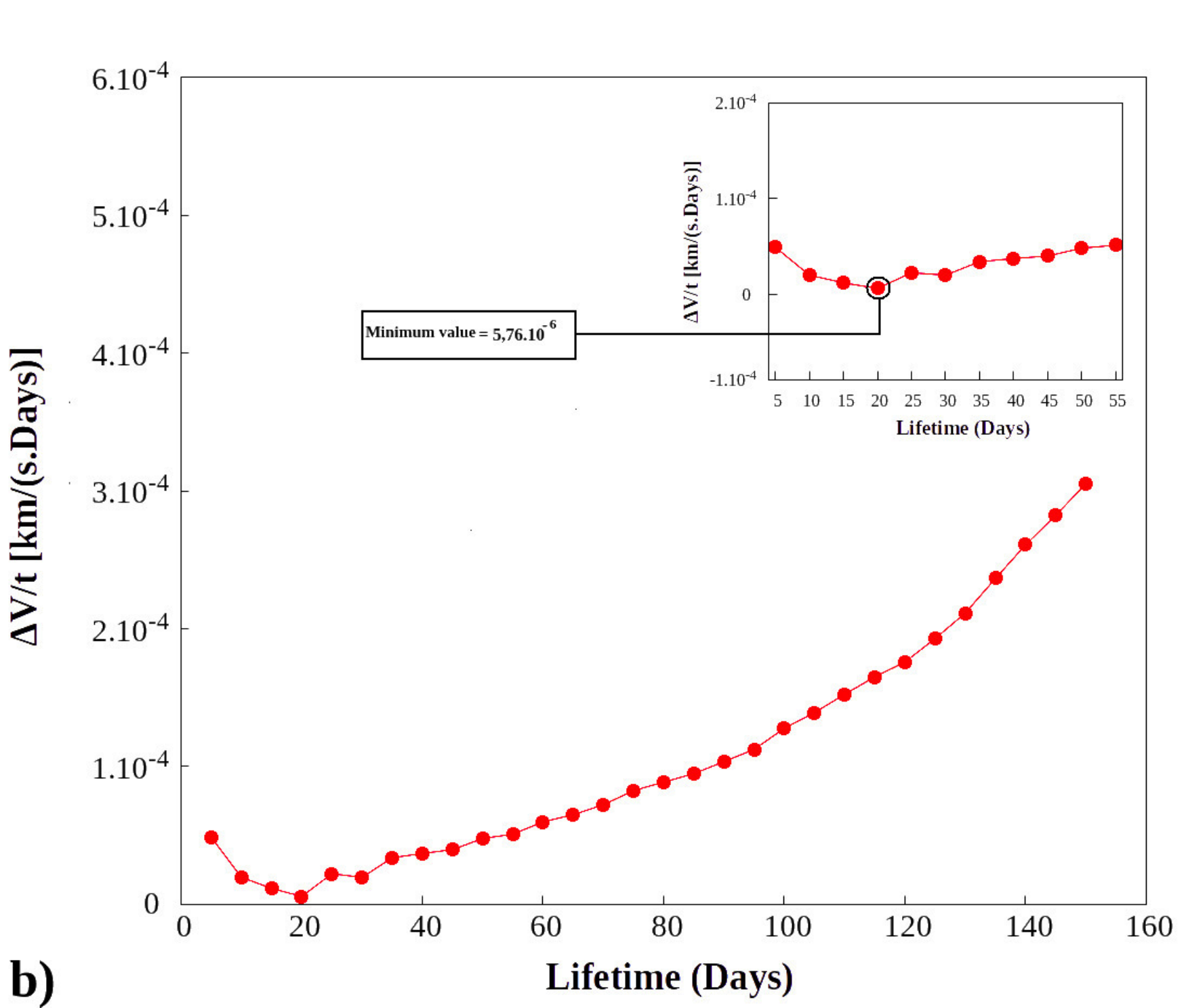}}
 \caption{Fuel economy analysis of an orbital maneuver through the $(\Delta V/t)$ versus lifetime relationship. Minimum value of $(\Delta V/t)$ for all analyzed cases. a) Case studied for $e = 10^{-4}$. b) Case studied for $e = 10^{-2}$. Initial conditions: $a = 1000$~km, $I = 90^\circ$, $\omega = 0^\circ$ and $\Omega = 0^\circ$.}   
\label{fig:deltavt}
\end{figure}

Making two  maneuvers $m_P$ will give a $\Delta V$ of $5.48\times10^{-2}$~km/s and a little over 100~days of life. Making two maneuvers $m_3$ would give a $\Delta V$ of $2.38\times10^{-3}$~km/s for $\approx 23$~days. The cost of the maneuver $m_P$ is much higher than $m_3$, since the $\Delta V$ used in the maneuver with a few days of life is 23 times smaller. Making another analysis of $m_P$ with another maneuver, now $m_2$, we confirm that, for this case, $m_3$ is also more economical. To perform two maneuvers $m_2$, the $\Delta V$ found is $\approx 1.962\times10^{-2}$~km/s to achieve a lifetime slightly greater than 60~days. This value is eight times greater than the $\Delta V$ of two maneuvers $m_3$. Thus, it is concluded that the best maneuver for $e=10^{-1}$ is $m_3$.

The results discussed in this section were essential to show the best conditions for performing maneuvers involving two coplanar elliptical orbits. Our simulations showed that the most economical maneuvers are those performed with little variations of the orbital elements. In the figures where the relationship between the variations of these elements and the maneuver time is presented, it is evident that these maneuvers are the most viable.

Another essential fact taken from the results presented in this section is that the lower-cost maneuvers are those for final orbits with eccentricity equal to $10^{-2}$ and $10^{-1}$. The maneuvers performed to reach a final orbit with these values of $e$ showed low values of $\Delta V$ compared to the other maneuvers performed for final orbits with $e$  equal to $10^{-4} $ and $10^{-3}$. The maneuvers whose final destination was an orbit with $e=10^{-1}$ showed a value of $\Delta V$ up to 95\% lower than the maneuvers performed for the other values of $e$ analyzed. Furthermore, it is evident that the maneuvers for final orbits with $e=10^{-3}$ and $e=10^{-4}$ have very close $\Delta V$. Thus, it is more advantageous to carry out the transfer to a final orbit with $e=10^{-3}$ since the lifetime for an orbit with this eccentricity is greater, about 45~days more for an orbit with $e=10^{-4}$.
All parameters used in performing the maneuvers discussed in this section are shown in Table \ref{tab:man_4}.

\begin{table}[ht]
\caption{Values of $\Delta V$ and initial conditions used in carrying out the maneuvers $m_P$, $m_1$, $m_2$ and $m_3$ for final values of eccentricity $e=10^{-1}$ and $m_P$, $m_1$, $m_2$, $m_3$, $m_4$ and $m_5$ for $e=10^{-2}$, $e=10^{-3}$ and $e=10^{-4}$. $\mathbf{T_{m}}$ is the time the maneuver was performed and $\mathbf{T_{c}}$ is the collision time.}
\label{tab:man_4}
\scalebox{0.46}{
\begin{tabular}{@{}lllllllll@{}}
   \hline
$\mathbf{e_f}$  & $\mathbf{\Delta V}$ \textbf{(km/s)} & $\mathbf{a_i}$ \textbf{(km)} & $\mathbf{e_i}$ & $\mathbf{I_i (^\circ)}$ & $\mathbf{r_{pi}}$ \textbf{(km)}  & $\mathbf{r_{api}}$ \textbf{(km)} & $\mathbf{T_{m}}$ \textbf{(Days)} &  $\mathbf{T_{c}}$ \textbf{(Days)} \\ 
\hline
\multicolumn{1}{c}{\multirow{4}{*}{$10^{-1}$}} & $ 2.74\times10^{-2}$ (m$_P$) & $ 9.998\times10^{2}$  & $2.098\times10^{-1}$  &  89.5 & 790.0 & 1209.5 & 50.83  & \multicolumn{1}{c}{\multirow{4}{*}{52}} \\ 

\multicolumn{1}{c}{} & $1.84\times10^{-2}$ (m$_1$) & $9.994\times10^{2}$  & $1.744\times10^{-1}$  & 89.5  & 825.07 & 1173.7 & 42.12 & \multicolumn{1}{c}{\multirow{3}{*}{}} \\   

\multicolumn{1}{c}{} & $9.81\times10^{-3}$ (m$_2$) &  $1.0\times10^{3}$ & $1.399\times10^{-1}$  & 89.9  & 860.0 & 1140.0 & 30.46 &   \multicolumn{1}{c}{\multirow{3}{*}{}} \\ 

\multicolumn{1}{c}{} & $1.19\times10^{-3}$ (m$_3$) & $ 9.998\times10^{2}$  & $1.048\times10^{-1}$  & 89.5  & 895.0 & 1104.0 & 11.53  &  \multicolumn{1}{c}{\multirow{3}{*}{}}  \\ 
\hline
\multicolumn{1}{c}{\multirow{6}{*}{$10^{-2}$}} & $4.99\times10^{-2}$ (m$_P$) & $9.988\times10^{2}$ & $2.089\times10^{-1}$  & 89.4  & 790.0 & 1207.0 & 152.68  & \multicolumn{1}{c}{\multirow{6}{*}{153. 2}} \\ 

\multicolumn{1}{c}{} & $4.10\times10^{-2}$ (m$_1$) & $9.995\times10^{2}$  & $1.745\times10^{-1}$  & 88.9 & 825.01   & 1174.0 & 145.9 & \multicolumn{1}{c}{\multirow{6}{*}{}} \\ 

\multicolumn{1}{c}{} & $2.92\times10^{-2}$ (m$_2$) & $9.991\times10^{2}$  & $1.389\times10^{-1}$  & 89.6 & 860.26   & 1137.90 & 135.0  &   \multicolumn{1}{c}{\multirow{6}{*}{}} \\   

\multicolumn{1}{c}{} & $2.31\times10^{-2}$ (m$_3$) & $9.989\times10^{2}$  & $1.039\times10^{-1}$  & 89.5 & 895.07   & 1102.80 & 122.35 &  \multicolumn{1}{c}{\multirow{6}{*}{}}  \\ 

\multicolumn{1}{c}{} & $1.45\times10^{-2}$ (m$_4$) & $9.993\times10^{2}$  & $6.931\times10^{-2}$  & 89.7 & 930.07   & 1068.60 & 104.85 &  \multicolumn{1}{c}{\multirow{6}{*}{}}  \\   

\multicolumn{1}{c}{} & $6.04\times10^{-3}$ (m$_5$) & $9.999\times10^{2}$  & $3.482\times10^{-2}$  & 89.8 & 965.14   & 1034.70 & 74.110 &  \multicolumn{1}{c}{\multirow{6}{*}{}}  \\ 

\hline 
\multicolumn{1}{c}{\multirow{6}{*}{$10^{-3}$}} & $5.23\times10^{-2}$ (m$_P$) & $9.988\times10^{2}$  & $2.089\times10^{-1}$  & 87.6  & 790.15 & 1207.56 & 312.57  & \multicolumn{1}{c}{\multirow{6}{*}{313}} \\ 

\multicolumn{1}{c}{} & $4.33\times10^{-2}$ (m$_1$) & $9.999\times10^{2}$  & $1.747\times10^{-1}$  & 87.9  & 825.18 & 1174.67 & 304.44 & \multicolumn{1}{c}{\multirow{3}{*}{}} \\   

\multicolumn{1}{c}{} & $3.43\times10^{-2}$ (m$_2$) & $9.996\times10^{2}$  & $1.396\times10^{-1}$  & 88.0  & 860.10 & 1139.22 & 295.15  &   \multicolumn{1}{c}{\multirow{3}{*}{}} \\  

\multicolumn{1}{c}{} & $2.54\times10^{-2}$ (m$_3$) & $9.99\times10^{2}$  & $1.041\times10^{-1}$  & 88.2  & 895.0 & 1103.01 & 282.04  &  \multicolumn{1}{c}{\multirow{3}{*}{}}  \\  

\multicolumn{1}{c}{} & $1.68\times10^{-2}$ (m$_4$) & $9.998\times10^{2}$  & $6.986\times10^{-2}$  & 88.6  & 930.03 & 1069.74 & 264.93 &  \multicolumn{1}{c}{\multirow{3}{*}{}}  \\ 

\multicolumn{1}{c}{} & $8.01\times10^{-3}$ (m$_5$) & $9.989\times10^{2}$  & $3.388\times10^{-2}$  & 88.9  & 965.05 & 1032.76 & 234.19 &  \multicolumn{1}{c}{\multirow{3}{*}{}}  \\ 
\hline
\multicolumn{1}{c}{\multirow{6}{*}{$10^{-4}$}} & $5.26\times10^{-2}$ (m$_P$) & $9.993\times10^{2}$  & $2.094\times10^{-1}$  & 87.7  & 790.0 & 1208.59 & 267.26 & \multicolumn{1}{c}{\multirow{6}{*}{268.536}} \\ 

\multicolumn{1}{c}{} & $4.34\times10^{-2}$ (m$_1$) & $9.989\times10^{2}$  & $1.74\times10^{-1}$  & 88.3  & 825.09 & 1172.88 & 259.87 & \multicolumn{1}{c}{\multirow{3}{*}{}} \\ 

\multicolumn{1}{c}{} & $3.45\times10^{-2}$ (m$_2$) & $9.991\times10^{2}$  & $1.392\times10^{-1}$  & 88.5  & 860.01 & 1138.35 & 248.59  &   \multicolumn{1}{c}{\multirow{3}{*}{}} \\   

\multicolumn{1}{c}{} & $2.56\times10^{-2}$ (m$_3$) & $9.988\times10^{2}$  & $9.97\times10^{-2}$  & 88.9  & 895.0 & 1103.21 & 234.86  &  \multicolumn{1}{c}{\multirow{3}{*}{}}  \\ 

\multicolumn{1}{c}{} & $1.7\times10^{-2}$ (m$_4$) & $9.996\times10^{2}$  & $6.951\times10^{-2}$  & 88.8  & 930.16 & 1069.15 & 218.20  &  \multicolumn{1}{c}{\multirow{3}{*}{}}  \\  

\multicolumn{1}{c}{} & $8.31\times10^{-3}$ (m$_5$) & $9.991\times10^{2}$  & $3.42\times10^{-2}$  & 89.0  & 965.01  & 1033.37 & 187.40  &  \multicolumn{1}{c}{\multirow{3}{*}{}}  \\
\hline 
\end{tabular}}
\end{table}

Another way of showing how economical a maneuver is concerning the others is to evaluate the $\Delta V$ spent per day to perform the maneuver $(\Delta V/t)$ as a function of the lifetime. This analysis is shown in previous works such as \cite{Ferreira2022} to find better points for carrying out orbital maneuvers of vehicles around natural satellites. For example, figure \ref{fig:deltavt} shows this relationship as a function of a lifetime for some values of the eccentricities.   

Figure \ref{fig:deltavt} shows the amount of fuel per day that a maneuver requires to correct the position of an orbit. The figure also presents a minimum value of $\Delta V/t$; however, it is necessary to analyze the feasibility of this value as a function of the number of days of lifetime. 

In Figure \ref{fig:deltavt}a, where $e=10^{-4}$, the best values are found in a range of 20-100~days. In 25~days, a minimum value of $2.47\times10^{-6}$~km/(s.day) is found. At this point, the $\Delta V$ is $6.19\times10^{-5}$~km/s. With 12 maneuvers like this, we already achieve the same lifetime as the maneuver $m_P$ performed one day before the collision and with a much smaller minimum fuel expenditure.

In this case, for $e=10^{-3}$, the orbital duration is longer, causing the variation of orbital elements to be slower. Thus, the minimum $\Delta V/t$ value is found in a range of 25-150~days, more precisely in 95~days. This value of $1.37\times10^{-7}$~km/(s.day) is equivalent to a $\Delta V$ of $1.31\times10^{-5}$~km/s, as in the previous case, which is the best $\Delta V$ in terms of lifetime. Only a few maneuvers with this value are necessary to increase in lifetime significantly. 

In the analysis presented in Figure \ref{fig:deltavt}b, for $e=10^{-2}$, the smallest values of $\Delta V/t$ are seen for a few days of integration. Due to the high eccentricity value, the minimum values of $\Delta V/t$ are around 5 to 40~days. In 20~days, we have the smallest value, $5.76\times10^{-6}$~km/(s.day). At this point, $\Delta V$ is $1.15\times10^{-4}$~km/s. Comparing the $\Delta V$ found for the maneuver $m_P$ made days before the collision and the $\Delta V$ related to the highlighted point in Figure \ref{fig:deltavt}b, it is much more feasible to make several maneuvers with the minimum value than to perform the maneuver days before the collision. Making eight maneuvers of $1.15\times10^{-4}$~km/s, we achieve the orbital duration of the maneuver $m_P$, using much less fuel.

 For $e = 10^{-1}$ in this figure, it is possible to note that $\Delta V/t$ behaves similarly to those presented in the previous cases. The minimum value found is $1.82\times10^{-5}$~km/(s.day), in five days of integration. The related $\Delta V$ is $9.11\times10^{-5}$~km/s, and in this case, it is also the best $\Delta V$ value for this case.

\section{Maneuvers to correct the periapsis argument $\omega$}
\label{sec:peri}
Recently, \cite{Xavier2022, Ferreira2022, Thamis2022} presented a study showing that values different from zero for the argument of periapsis ($\omega$) and the longitude of the ascending node ($\Omega$) are essential to increase the lifetime of an orbit. These works also present values of ($\omega$) and ($\Omega$) capable of prolonging the orbital lifetime. In \cite{Cinelli2019, Zhang2019}, maneuver models for correcting changes in the argument of periapsis are proposed. These maneuvers are essential because they ensure that this angle remains close to the values found in the regions of the most extended orbital lifetime and thus prolong the duration of the orbit.

Depending on the initial parameters considered for launching a spacecraft, the argument of periapsis may not be located in the desired geographic position, and therefore corrections become necessary.

\begin{figure}[h]
\centering
    \fbox{\includegraphics[width=5.5cm]{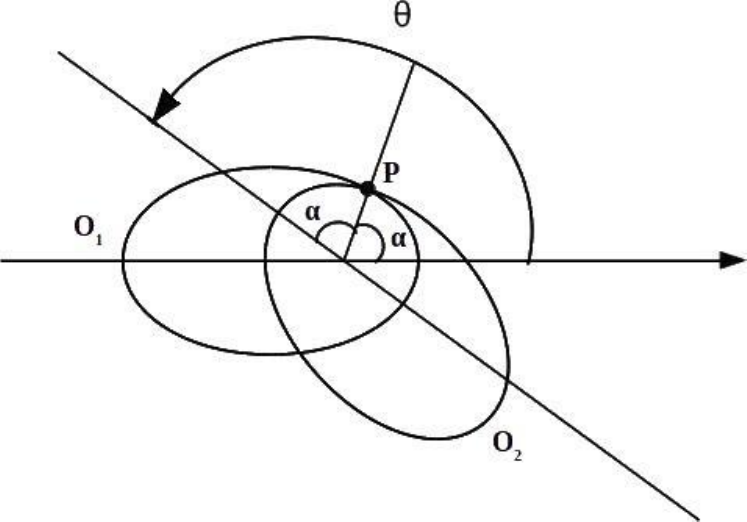}}
    \caption{Rotation of an elliptical orbit by an angle $\theta$. $\theta$ is also the $\omega$ value to be corrected in the maneuver.}
    \label{fig:elip_rot}
\end{figure}

Figure \ref{fig:elip_rot} shows the rotation of an ellipse by an angle $\theta$, which also represents the variation of $\omega$.

If it is desired to change the periapsis argument by a value $\theta$ without changing the semi-major axis $a$ and the eccentricity $e$, the maneuver must be performed in the plane of the orbit. The realization of a maneuver like this is given by \cite{Sidi1997}:
\begin{equation}
   \Delta V= 2\sqrt{\dfrac{\mu}{a(1-e^2)}} e \sin{\left(\dfrac{\theta}{2}\right)}
  \label{eq:peri_2}
\end{equation}

\noindent where $a$ is the semi-major axis, $e$ is the eccentricity, $\theta$ is the desired change in the argument of periapsis, $P$ is the common point of the two orbits, the initial orbit and the orbit to be occupied after the maneuver, and $\Delta V$ is the total velocity to be added to the orbit to correct $\omega$. However, it is important to note that equation \ref{eq:peri_2} is only valid for constant values of $a$ and $e$, which becomes ineffective for high values of these parameters.

Given the above, values of $\Delta V$ will be calculated for some cases of variation in the argument of periapsis. To effectively use equation \ref{eq:peri_2}, $\Delta V$ will be calculated over a time interval where variations in the semi-major axis and eccentricity are tiny.  
For the cases analyzed in this work, the best time to perform the periapsis argument correction maneuver would be within 50 days, after which the periapsis argument will have undergone significant changes, as well as the eccentricity \cite{Cinelli2019}. Therefore, $\Delta V$ will be calculated at a point before or within 50 days and the final desired orbits will have the same eccentricity values analyzed in the previous section ($e_f=10^{-4}, 10^{-3}, 10^{-2}, 10^{-1})$.

For an orbit with a final eccentricity of $10^{-4}$, $a=1000$~km, $I=90^\circ$, and $\omega=\Omega=0^\circ$, the semi-major axis suffers a small variation during the whole integration, around 1 km. However, the eccentricity increases around 80 days, and then the maneuver will be performed in 50 days. When the orbit reaches this lifetime, its parameters are the semi-major axis $9.997 \times 10^{2}$~km, $e=9.7 \times 10^{-4}$, $\theta=2.32 \times 10^{2} $ degrees (variation of periapsis argument). Considering these parameters, a $\Delta V$ of $1.92 \times 10^{-3}$ km/s is obtained.

For $e_f=10^{-3}$, the other initial conditions are the same as for $e=10^{-4}$. In this case, the semi-major axis also has a slight variation, around 1 km. Although the eccentricity remains very low until around 100 days of integration, the periapsis argument has a high  variation, so the maneuver will be performed at 50 days. At 50 days, the orbit has a semi-major axis of $9.99 \times 10^{2}$ km, an eccentricity of $8.6 \times 10^{-4}$, and $\theta$ equals 8.0 degrees. Therefore, the $\Delta V$ for this point in the orbit was $5.85 \times 10^{-5}$ km/s. In the previous section, the results indicated that the most viable maneuver would be for an orbit with $e$ equal to $10^{-3}$, since orbits with this eccentricity have a longer useful life and remain with their elements with low variation for a long period. This is also shown here, as $\Delta V=5.85 \times 10^{-5}$ km/s is a considerable value for such a maneuver.

The best interval for performing the maneuver for $e_f=10^{-2}$ is up to 20 days. In this time, the orbit has a semi-major axis of $9.99 \times 10^{2}$ km, an eccentricity of $1.23 \times 10^{-2}$, $\theta$ equal to 22 degrees, and thus, $\Delta V =2.29 \times 10^{-3}$ km/s. The initial parameters are the same as those used in the previous cases. 

In the case of $e_f=10^{-1}$, since the initial eccentricity is very high, the best interval for maneuver execution is up to 8 days. The semi-major axis and the eccentricity within this interval show little variations from their initial values, remaining almost constant. Thus, any point within this interval can have its values used in equation \ref{eq:peri_2} to calculate the desired $\Delta V$. For example, with eight days of integration, the semi-major axis is $9.99 \times 10^{2}$ km, the eccentricity is $1.0 \times 10^{-1}$, $\theta=1.8^\circ$, and thus the $\Delta V$ required is $6.68 \times 10^{-3}$ km/s.

The results presented in this section demonstrate a maneuver's feasibility to correct the periapsis argument. The $\Delta V_s$ found for different eccentricities suggest that the most viable maneuver is for an orbit with $e$ equal to $10^{-3}$, since the value found for this case is of the order of $10^{-5}$ km/s. In this case, it would require approximately $0.058$ m/s to correct only $1^\circ$ of variation in the argument of periapsis. 

In addition, we have shown that, as presented in the previous section, maneuvers are more effective when performed with a short duration of the orbit. 

A summary of the entire discussion in this section is presented in Table \ref{tab:peri}.
\begin{table}[ht]
\caption{Values of $\Delta V$ and initial conditions used in carrying out the maneuvers to correct the variation of the periapsis argument $(\theta=\omega)$. Where $T_{m}$ is the time to perform the maneuver and $e_f$ is the exectricity of the intended final orbit.}
\label{tab:peri}
\scalebox{0.65}{
\begin{tabular}{@{}lllllll@{}}
   \hline
$\mathbf{e_f}$  & $\mathbf{\Delta V}$ \textbf{(km/s)} & $\mathbf{a_i}$ \textbf{(km)} & $\mathbf{e_i}$ &  $\mathbf{\theta=\omega (^\circ)}$ &  $\mathbf{T_{m}}$ \textbf{(Days)}  \\ 
\hline
\multicolumn{1}{c}{\multirow{1}{*}{$e=10^{-1}$}} & $ 6.68 \times 10^{-3}$  & $9.99 \times 10^2  $  & $1.0 \times 10^{-1}$  &  1.8 & 8  \\ \hline

\multicolumn{1}{c}{\multirow{1}{*}{$e=10^{-2}$}} & $ 2.29 \times 10^{-3}$  & $9.99 \times 10^2  $  & $1.23 \times 10^{-2}$  &  22  & 20  \\ \hline

\multicolumn{1}{c}{\multirow{1}{*}{$e=10^{-3}$}} & $ 5.85 \times 10^{-5}$  & $9.99 \times 10^2  $  & $8.6 \times 10^{-4}$  &  8  & 50  \\ \hline

\multicolumn{1}{c}{\multirow{1}{*}{$e=10^{-4}$}} & $ 1.92 \times 10^{-3}$  & $9.997 \times 10^2  $  & $9.7 \times 10^{-4}$  &  232    & 50  \\ \hline 
\end{tabular}}
\end{table}
 
\begin{figure}[!h]
    \centering
    \fbox{\includegraphics[width = 6.0 cm]{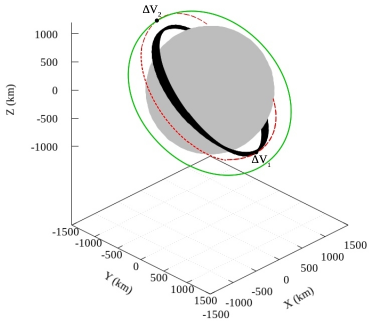}}
    \caption{Representation of a maneuver using a minimum $\Delta V$ equal to $8.01 \times 10^{-3}$. Initial conditions: $a=9.989 \times 
    10^2$ km, $e_i=3.388 \times 10^{-2}$, $I_i=88.9^\circ$.}
    \label{fig:manobra}
\end{figure}

In Figure \ref{fig:manobra} we provide a representation of a maneuver with one of the best values of $\Delta V$, its initial orbit (black line), transfer ellipse (red line) and the desired final elliptical orbit (green). Values for performing this maneuver are provided in Table \ref{tab:man_4} and in the description of the figure.
 \section{Final Comments}
\label{sec:conclusoes}

In this work, we present a study of maneuvers for high inclination. To investigate which orbit gives a more significant variation over its orbital duration, we consider four initial inclinations $60^{\circ}$, $70^{\circ}$, $80^{\circ}$ and $90^{\circ}$, with final orbit radii between 1000-2000~km.

Our results showed that orbits with an initial inclination of $70^\circ$ presented a greater deflection than the other inclinations. We also observed that the best maneuvers were found for smaller orbit radii, between 1000-1050~km. We found a minimum $\Delta V$ of $5.4\times10^{-05}$~km/s for all inclinations.

We also presented a set of simulations for maneuvers between two coplanar elliptical orbits, with final eccentricities of $10^{-4}$, $10^{-3}$, $10^{-2}$, and $10^{-1}$. We consider a few points in the orbit to compare the feasibility of one maneuver another, with six points for the values of $e=10^{-4}$, $e=10^{-3}$, and $e=10^{-2}$, and three points for $e=10^{-1}$.

This set of simulations showed that maneuvers carried out within a few days of life were more economical than those carried out near collision. These results are confirmed by analyzing the $\Delta V/t$ ratio, which showed that the velocity per maneuver time was indeed lower for orbits with a few days of duration and, consequently, with a lower variation of the orbital elements.

Finally, we present an option for a maneuver capable of correcting the argument of periapsis variation. We calculate the $\Delta V$ at the points of interest. Our results corroborate the previous results showing that maneuvers carried out within a few days of integration are more feasible than those carried out near collision.

\bmhead{Acknowledgments}

The authors thank Improvement Coordination Higher Education Personnel - Brazil (CAPES) - Financing Code 001. The Center for Mathematical Sciences Applied to Industry (Ce-MEAI),  FAPESP (Proc~2013/07375-0 and Proc~2016/23542-1). SMGW also thanks CNPq (Proc~313043/2020-5) for the financial support. ABAP thanks CNPq (Proc~309089/2021-2). This publication has been supported by the RUDN University Scientific Projects Grant System, project N$^\circ$~202235-2-000. 

\bmhead{Availability of data} All data generated or analyzed during this study are included in this published article in the form of figures.

\bibliography{sn-bibliography}


\begin{thebibliography}{14}
\ifx \bisbn   \undefined \def \bisbn  #1{ISBN #1}\fi
\ifx \binits  \undefined \def \binits#1{#1}\fi
\ifx \bauthor  \undefined \def \bauthor#1{#1}\fi
\ifx \batitle  \undefined \def \batitle#1{#1}\fi
\ifx \bjtitle  \undefined \def \bjtitle#1{#1}\fi
\ifx \bvolume  \undefined \def \bvolume#1{\textbf{#1}}\fi
\ifx \byear  \undefined \def \byear#1{#1}\fi
\ifx \bissue  \undefined \def \bissue#1{#1}\fi
\ifx \bfpage  \undefined \def \bfpage#1{#1}\fi
\ifx \blpage  \undefined \def \blpage #1{#1}\fi
\ifx \burl  \undefined \def \burl#1{\textsf{#1}}\fi
\ifx \doiurl  \undefined \def \doiurl#1{\url{https://doi.org/#1}}\fi
\ifx \betal  \undefined \def \betal{\textit{et al.}}\fi
\ifx \binstitute  \undefined \def \binstitute#1{#1}\fi
\ifx \binstitutionaled  \undefined \def \binstitutionaled#1{#1}\fi
\ifx \bctitle  \undefined \def \bctitle#1{#1}\fi
\ifx \beditor  \undefined \def \beditor#1{#1}\fi
\ifx \bpublisher  \undefined \def \bpublisher#1{#1}\fi
\ifx \bbtitle  \undefined \def \bbtitle#1{#1}\fi
\ifx \bedition  \undefined \def \bedition#1{#1}\fi
\ifx \bseriesno  \undefined \def \bseriesno#1{#1}\fi
\ifx \blocation  \undefined \def \blocation#1{#1}\fi
\ifx \bsertitle  \undefined \def \bsertitle#1{#1}\fi
\ifx \bsnm \undefined \def \bsnm#1{#1}\fi
\ifx \bsuffix \undefined \def \bsuffix#1{#1}\fi
\ifx \bparticle \undefined \def \bparticle#1{#1}\fi
\ifx \barticle \undefined \def \barticle#1{#1}\fi
\bibcommenthead
\ifx \bconfdate \undefined \def \bconfdate #1{#1}\fi
\ifx \botherref \undefined \def \botherref #1{#1}\fi
\ifx \url \undefined \def \url#1{\textsf{#1}}\fi
\ifx \bchapter \undefined \def \bchapter#1{#1}\fi
\ifx \bbook \undefined \def \bbook#1{#1}\fi
\ifx \bcomment \undefined \def \bcomment#1{#1}\fi
\ifx \oauthor \undefined \def \oauthor#1{#1}\fi
\ifx \citeauthoryear \undefined \def \citeauthoryear#1{#1}\fi
\ifx \endbibitem  \undefined \def \endbibitem {}\fi
\ifx \bconflocation  \undefined \def \bconflocation#1{#1}\fi
\ifx \arxivurl  \undefined \def \arxivurl#1{\textsf{#1}}\fi
\csname PreBibitemsHook\endcsname

\bibitem{bib1}
\begin{barticle}
\bauthor{\bsnm{Hofstadter}, \binits{M.}},
\bauthor{\bsnm{Simon}, \binits{A.}},
\bauthor{\bsnm{Atreya}, \binits{S.}},
\bauthor{\bsnm{Banfield}, \binits{D.}},
\bauthor{\bsnm{Fortney}, \binits{J.J.}},
\bauthor{\bsnm{Hayes}, \binits{A.}},
\bauthor{\bsnm{Hedman}, \binits{M.}},
\bauthor{\bsnm{Hospodarsky}, \binits{G.}},
\bauthor{\bsnm{Mandt}, \binits{K.}},
\bauthor{\bsnm{Masters}, \binits{A.}},
\bauthor{\bsnm{Showalter}, \binits{M.}},
\bauthor{\bsnm{Soderlund}, \binits{K.M.}},
\bauthor{\bsnm{Turrini}, \binits{D.}},
\bauthor{\bsnm{Turtle}, \binits{E.}},
\bauthor{\bsnm{Reh}, \binits{K.}},
\bauthor{\bsnm{Elliott}, \binits{J.}},
\bauthor{\bsnm{Arora}, \binits{N.}},
\bauthor{\bsnm{Petropoulos}, \binits{A.}}:
\batitle{Uranus and neptune missions: A study in advance of the next planetary
  science decadal survey}.
\bjtitle{Planetary and Space Science}
\bvolume{177},
\bfpage{104680}
(\byear{2019}).
\doiurl{10.1016/j.pss.2019.06.004}
\end{barticle}
\endbibitem

\bibitem{Grasset2013}
\begin{barticle}
\bauthor{\bsnm{Grasset}, \binits{O.}},
\bauthor{\bsnm{Dougherty}, \binits{M.K.}},
\bauthor{\bsnm{Coustenis}, \binits{A.}},
\bauthor{\bsnm{Bunce}, \binits{E.}},
\bauthor{\bsnm{Erd}, \binits{C.}},
\bauthor{\bsnm{Titov}, \binits{D.V.}},
\bauthor{\bsnm{Blanc}, \binits{M.}},
\bauthor{\bsnm{Coates}, \binits{A.}},
\bauthor{\bsnm{Drossart}, \binits{P.}},
\bauthor{\bsnm{Fletcher}, \binits{L.}},
\bauthor{\bsnm{Hussmann}, \binits{H.}},
\bauthor{\bsnm{Jaumann}, \binits{R.}},
\bauthor{\bsnm{Krupp}, \binits{N.}},
\bauthor{\bsnm{Lebreton}, \binits{J.-P.}},
\bauthor{\bsnm{Prieto-Ballesteros}, \binits{O.}},
\bauthor{\bsnm{Tortora}, \binits{P.}},
\bauthor{\bsnm{Tosi}, \binits{F.}},
\bauthor{\bsnm{Van~Hoolst}, \binits{T.}}:
\batitle{Jupiter icy moons explorer (juice): An esa mission to orbit ganymede
  and to characterise the jupiter system}.
\bjtitle{Planetary and Space Science}
\bvolume{78},
\bfpage{1}--\blpage{21}
(\byear{2013}).
\doiurl{10.1016/j.pss.2012.12.002}
\end{barticle}
\endbibitem

\bibitem{Xavier2022}
\begin{botherref}
\oauthor{\bsnm{Xavier}, \binits{J.}},
\oauthor{\bsnm{Prado}, \binits{A.}},
\oauthor{\bsnm{Winter}, \binits{S.}},
\oauthor{\bsnm{Amarante}, \binits{A.}}:
Mapping long-term natural orbits about titania, a satellite of uranus
(2022)
\end{botherref}
\endbibitem

\bibitem{Thamis2022}
\begin{barticle}
\bauthor{\bsnm{{Ferreira}}, \binits{T.C.F.C.}},
\bauthor{\bsnm{{Prado}}, \binits{A.F.B.A.}},
\bauthor{\bsnm{{Giuliatti Winter}}, \binits{S.M.}},
\bauthor{\bsnm{{Ferreira}}, \binits{L.S.}}:
\batitle{{Mapping Natural Orbits around Io}}.
\bjtitle{Symmetry}
\bvolume{14}(\bissue{7}),
\bfpage{1478}
(\byear{2022}).
\doiurl{10.3390/sym14071478}
\end{barticle}
\endbibitem

\bibitem{Cinelli2022}
\begin{botherref}
\oauthor{\bsnm{Cinelli}, \binits{M.}}:
Probe lifetime around natural satellites with obliquity.
Astrodynamics
\textbf{6}
(2022).
\doiurl{10.1007/s42064-022-0145-1}
\end{botherref}
\endbibitem

\bibitem{Cinelli2019}
\begin{botherref}
\oauthor{\bsnm{Cinelli}, \binits{M.}},
\oauthor{\bsnm{Ortore}, \binits{E.}},
\oauthor{\bsnm{Circi}, \binits{C.}}:
Long lifetime orbits for the observation of europa.
Journal of Guidance, Control, and Dynamics
(2019)
\end{botherref}
\endbibitem

\bibitem{Ferreira2022}
\begin{barticle}
\bauthor{\bsnm{{Ferreira}}, \binits{L.S.}},
\bauthor{\bsnm{{Sfair}}, \binits{R.}},
\bauthor{\bsnm{{Prado}}, \binits{A.F.B.A.}}:
\batitle{{Lifetime and Dynamics of Natural Orbits around Titan}}.
\bjtitle{Symmetry}
\bvolume{14}(\bissue{6}),
\bfpage{1243}
(\byear{2022}).
\doiurl{10.3390/sym14061243}
\end{barticle}
\endbibitem

\bibitem{Chen2014}
\begin{barticle}
\bauthor{\bsnm{{Chen}}, \binits{E.M.A.}},
\bauthor{\bsnm{{Nimmo}}, \binits{F.}},
\bauthor{\bsnm{{Glatzmaier}}, \binits{G.A.}}:
\batitle{{Tidal heating in icy satellite oceans}}.
\bjtitle{Icarus}
\bvolume{229},
\bfpage{11}--\blpage{30}
(\byear{2014}).
\doiurl{10.1016/j.icarus.2013.10.024}
\end{barticle}
\endbibitem

\bibitem{Tzirt2009}
\begin{barticle}
\bauthor{\bsnm{{Tzirti}}, \binits{S.}},
\bauthor{\bsnm{{Tsiganis}}, \binits{K.}},
\bauthor{\bsnm{{Varvoglis}}, \binits{H.}}:
\batitle{{Quasi-critical orbits for artificial lunar satellites}}.
\bjtitle{Celestial Mechanics and Dynamical Astronomy}
\bvolume{104}(\bissue{3}),
\bfpage{227}--\blpage{239}
(\byear{2009}).
\doiurl{10.1007/s10569-009-9207-4}
\end{barticle}
\endbibitem

\bibitem{Tzirt2010}
\begin{barticle}
\bauthor{\bsnm{{Tzirti}}, \binits{S.}},
\bauthor{\bsnm{{Tsiganis}}, \binits{K.}},
\bauthor{\bsnm{{Varvoglis}}, \binits{H.}}:
\batitle{{Effect of 3rd-degree gravity harmonics and Earth perturbations on
  lunar artificial satellite orbits}}.
\bjtitle{Celestial Mechanics and Dynamical Astronomy}
\bvolume{108}(\bissue{4}),
\bfpage{389}--\blpage{404}
(\byear{2010}).
\doiurl{10.1007/s10569-010-9313-3}
\end{barticle}
\endbibitem

\bibitem{Scheeres2006}
\begin{barticle}
\bauthor{\bsnm{{Paskowitz}}, \binits{M.E.}},
\bauthor{\bsnm{{Scheeres}}, \binits{D.J.}}:
\batitle{{Design of Science Orbits About Planetary Satellites: Application to
  Europa}}.
\bjtitle{Journal of Guidance Control Dynamics}
\bvolume{29}(\bissue{5}),
\bfpage{1147}--\blpage{1158}
(\byear{2006}).
\doiurl{10.2514/1.19464}
\end{barticle}
\endbibitem

\bibitem{Chambers1999}
\begin{barticle}
\bauthor{\bsnm{{Chambers}}, \binits{J.E.}}:
\batitle{{A hybrid symplectic integrator that permits close encounters between
  massive bodies}}.
\bjtitle{MNRAS}
\bvolume{304}(\bissue{4}),
\bfpage{793}--\blpage{799}
(\byear{1999}).
\doiurl{10.1046/j.1365-8711.1999.02379.x}
\end{barticle}
\endbibitem

\bibitem{Zhang2019}
\begin{barticle}
\bauthor{\bsnm{Zhang}, \binits{G.}},
\bauthor{\bsnm{Zhang}, \binits{H.}},
\bauthor{\bsnm{Cao}, \binits{X.}}:
\batitle{New solutions to impulsive correction for argument of perigee using
  gauss’s variational equations}.
\bjtitle{Journal of Aerospace Engineering}
\bvolume{32},
\bfpage{04019071}
(\byear{2019}).
\doiurl{10.1061/(ASCE)AS.1943-5525.0001073}
\end{barticle}
\endbibitem

\bibitem{Sidi1997}
\begin{bbook}
\bauthor{\bsnm{Sidi}, \binits{M.J.}}:
\bbtitle{Spacecraft Dynamics and Control: A Practical Engineering Approach}.
\bpublisher{Cambridge University Press},
\blocation{Cambridge}
(\byear{1997})
\end{bbook}
\endbibitem

\end{thebibliography}


\end{document}